\pdfoutput=1

\documentclass[11pt,twoside,a4paper,cmspaper,final,collab]{cms-tdr}
\def\svnVersion{368093}\def\cmsCernNoTag{CERN-EP/2016-139}\def\cmsCernDate{\today}\def\cmsMessage{Published in Physical Review D as \href{http://dx.doi.org/10.1103/PhysRevD.94.052006}{\doi{10.1103/PhysRevD.94.052006.}}}

\begin{document}\cmsNoteHeader{TOP-12-042}

\hyphenation{had-ron-i-za-tion}
\hyphenation{cal-or-i-me-ter}
\hyphenation{de-vices}
\RCS$Revision: 364014 $
\RCS$HeadURL: svn+ssh://svn.cern.ch/reps/tdr2/papers/TOP-12-042/trunk/TOP-12-042.tex $
\RCS$Id: TOP-12-042.tex 364014 2016-08-08 21:07:47Z alverson $
\newlength\cmsFigWidth
\ifthenelse{\boolean{cms@external}}{\setlength\cmsFigWidth{0.96\columnwidth}}{\setlength\cmsFigWidth{0.48\textwidth}}
\ifthenelse{\boolean{cms@external}}{\providecommand{\cmsLeft}{top}}{\providecommand{\cmsLeft}{left}}
\ifthenelse{\boolean{cms@external}}{\providecommand{\cmsRight}{bottom}}{\providecommand{\cmsRight}{right}}
\ifthenelse{\boolean{cms@external}}{\providecommand{\NA}{\ensuremath{\cdots}\xspace}}{\providecommand{\NA}{\text{---}\xspace}}

\providecommand{\suppMaterial}{Tables \ref{tab:MET_xsections_7TeV_combined}--\ref{tab:WPT_xsections_8TeV_combined} of Appendix~\ref{s:additional_tables}}

\renewcommand{\tt}{\ttbar}

\newcommand{\met}{\MET}
\newcommand{\st}{\ensuremath{S_\mathrm{T}}\xspace}

\newcommand{\xsect}{cross section}
\newcommand{\xSect}{Cross section}
\newcommand{\rec}{\text{rec}}
\newcommand{\gen}{\text{gen}}

\ifthenelse{\boolean{cms@external}}{\newcommand{\wpt}{\ensuremath{ p_\mathrm{T}^{W} }\xspace}}{\newcommand{\wpt}{\ensuremath{ p_\mathrm{T}^{\mathrm{W}} }\xspace}}
\ifthenelse{\boolean{cms@external}}{\newcommand{\mtop}{\ensuremath{ m_{t} }\xspace}}{\newcommand{\mtop}{\ensuremath{ m_{\mathrm{t}} }\xspace}}
\ifthenelse{\boolean{cms@external}}{\newcommand{\pp}{\ensuremath{ pp }\xspace}}{\newcommand{\pp}{\ensuremath{ \mathrm{pp} }\xspace}}

\newcommand{\ST}{\st}
\newcommand{\MT}{\mt}

\newcommand{\pb}{\unit{pb}}
\newcommand{\sigmameasured}{\ensuremath{\sigma_{\mathrm{meas}}^{\mathrm{norm}}}}

\newcommand{\POWHEGPYTHIA}{{\textsc{powheg}\ +\ \textsc{pythia}}\xspace}
\newcommand{\POWHEGHERWIG}{{\textsc{powheg}\ +\ \textsc{herwig}}\xspace}
\newcommand{\MCATNLOHERWIG} {\textsc{mc@nlo}\ +\ \textsc{herwig}\xspace}

\newcommand{\MADGRAPHPYTHIA}{{\MADGRAPH\ +\ \textsc{pythia}}\xspace}

\newcommand{\POWHEGVONE}{\textsc{powheg v1}\xspace}
\newcommand{\POWHEGVTWO}{\textsc{powheg v2}\xspace}

\newcommand{\POWHEGVONEPYTHIA}{{\textsc{powheg v1}\ +\ \textsc{pythia}}\xspace}
\newcommand{\POWHEGVONEHERWIG}{{\textsc{powheg v1}\ +\ \textsc{herwig}}\xspace}
\newcommand{\POWHEGVTWOPYTHIA}{{\textsc{powheg v2}\ +\ \textsc{pythia}}\xspace}
\newcommand{\POWHEGVTWOHERWIG}{{\textsc{powheg v2}\ +\ \textsc{herwig}}\xspace}

\newcommand{\sqrtSeven}{\ensuremath{\sqrt{s}=7\TeV}\xspace}
\newcommand{\sqrtEight}{\ensuremath{\sqrt{s}=8\TeV}\xspace}
\newcommand{\sqrtSevenAndEight}{\ensuremath{\sqrt{s}=7\TeV\text{ and }\sqrt{s}=8\TeV}\xspace}

\newcommand{\x}{\ensuremath{\phantom{0}}}
\newcommand{\xx}{\ensuremath{\phantom{00}}}
\newcolumntype{.}{D{.}{.}{-1}}

\cmsNoteHeader{TOP-12-042}
\title{Measurement of the differential cross sections for top quark pair production as a function of kinematic event variables in \pp collisions at $\sqrt{s} = 7$ and 8\TeV}

\date{\today}

\abstract{
Measurements are reported of the normalized differential cross sections for top quark pair production with respect to four kinematic event variables: the missing transverse energy; the scalar sum of the jet transverse momentum ($p_{\mathrm{T}}$); the scalar sum of the \pt of all objects in the event; and the $p_{\mathrm{T}}$ of leptonically decaying $\PW$ bosons from top quark decays. The data sample, collected using the CMS detector at the LHC, consists of 5.0\fbinv of proton-proton collisions at $\sqrt{s} = 7\TeV$ and 19.7\fbinv at $\sqrt{s} = 8\TeV$. Top quark pair events containing one electron or muon are selected. The results are presented after correcting for detector effects to allow direct comparison with theoretical predictions. No significant deviations from the predictions of several standard model event simulation generators are observed.
}

\hypersetup{%
pdfauthor={CMS Collaboration},%
pdftitle={Measurement of differential top quark pair production cross sections as a function of kinematic event
variables in pp collisions at sqrt(s) = 7 TeV and 8 TeV},
pdfsubject={CMS},%
pdfkeywords={CMS, physics, top, met, semileptonic, unfolding}}

\maketitle

\section{Introduction}
\label{sec:introduction}

The CERN LHC produced millions of top quark
pairs (\tt) in 2011 and 2012. This allows for a detailed investigation of the kinematic event properties of
\ttbar production such as the missing transverse energy (\met),
the scalar sum of the jet transverse momenta (\HT),
the scalar sum of the transverse momenta
of all objects (\st), and
the transverse momentum (\wpt) of leptonically decaying $\PW$ bosons produced in top quark decays.
These measurements can be used to verify current theoretical models, along with their implementation
in simulations of \ttbar production, and also to measure rare standard model (SM)
processes such as \ttbar production in association with a  $\PW$, $\PZ$, or Higgs boson.
Since top quark pair production is a major background for many searches for physics beyond the SM,
it is important that the properties of \ttbar events are well understood.

Here, we report measurements carried out using the CMS detector~\cite{bib:JINST} at the LHC at
two different proton--proton center-of-mass energies. The data samples used include integrated
luminosities of 5.0\fbinv collected in 2011 at \sqrtSeven  and 19.7\fbinv from 2012 at \sqrtEight.
The \ttbar production cross section is measured as a function of \met, \HT, \st, and \wpt, corrected for detector effects,
and compared with the predictions from different event generators. Differential \ttbar cross sections
have previously been measured at the Tevatron~\cite{CDFDiffTop,DZeroDiffTop}, and at the
LHC~\cite{CMSDiffTop7TeV, CMSDiffTop8TeV, CMSDiffTopExtraJets8TeV, CMSDiffTopHighPt8TeV,ATLASDiffTop7TeV, ATLASDiffTopHighPt8TeV}.
These previous measurements study the \ttbar production cross section as a function of the top quark kinematics
and the kinematics of the \ttbar system. The results presented here are complementary,
since the \ttbar production cross section is measured as a function of variables that do not require the reconstruction
of the top quarks from their decay products.

Top quarks decay with close to 100\% probability into a $\PW$ boson and a
bottom quark. In this article, we consider the channel
in which one of the $\PW$ bosons decays leptonically into a charged lepton (electron or
muon) along with its associated neutrino, while the other $\PW$ boson decays hadronically.
This channel has a branching fraction of around 15\% for direct decay to each lepton
flavor and a relatively clean experimental signature, including an
isolated, high-transverse-momentum lepton, large \met\ from the undetected
neutrino, and multiple hadronic jets. Two jets are expected to contain $\PQb$ hadrons
from the hadronization of the $\PQb$ quarks produced directly in the
$\PQt \to \PQb \PW$ decay, while other jets (from the hadronic $\PW$ boson
decay or gluon radiation) will typically contain only light and charm quarks.

\section{The CMS detector}

The central feature of the CMS apparatus is a superconducting solenoid of 6\unit{m} internal diameter, providing a
magnetic field of 3.8\unit{T}. Within the solenoid volume are a silicon pixel and strip tracker, a lead
tungstate crystal electromagnetic calorimeter, and a brass and scintillator hadron calorimeter,
each composed of a barrel and two endcap sections. Muons are measured in gas-ionisation detectors embedded in the steel flux-return
yoke outside the solenoid. Extensive forward calorimetry complements the coverage provided by the barrel and endcap
detectors.

A more detailed description of the CMS detector, together with a definition of the coordinate system used and the
relevant kinematic variables, can be found in Ref.~\cite{bib:JINST}.

\section{Simulation}
\label{sec:mc_modelling}
For the Monte Carlo (MC) simulation of the \ttbar signal sample the leading-order \MADGRAPH v5.1.5.11 event generator~\cite{MadGraph}
is used with relevant matrix elements for up to three additional partons implemented.
Theoretical production \xsect\ values
of
$177.3\ ^{+4.6}_{-6.0}\ \mathrm{(scale)}\pm9.0\ \mathrm{(PDF +}\alpS \mathrm{)}\pb$ at \sqrtSeven,
and
$252.9\ ^{+6.4}_{-8.6}\ \mathrm{(scale)}\pm11.7\ \mathrm{(PDF +}\alpS \mathrm{)}\pb$ at \sqrtEight,
are used for the normalization of these samples.  These cross sections are calculated with the Top++2.0 program
to next-to-next-to-leading order (NNLO) in perturbative QCD, including soft-gluon resummation to
next-to-next-to-leading-logarithm (NNLL) order~\cite{Topplusplus}, and assuming a top quark mass $\mtop = 172.5\GeV$.
The first uncertainty comes from the independent variation of the renormalization ($\mu_{\mathrm{R}}$)
and factorization ($\mu_{\mathrm{F}}$) scales, while the second one is associated with variations in the
parton distribution function (PDF) and \alpS,
following the PDF4LHC prescription with the MSTW2008 68\% CL NNLO,
CT10 NNLO, and NNPDF2.3 5f FFN PDF sets~\cite{Botje:2011sn,Alekhin:2011sk,alphaSUnc,ct10TTbarXsec,pdfWithLHCData}.

The generated events are subsequently processed with \PYTHIA v6.426~\cite{pythia} for parton showering and hadronization.
The \PYTHIA parton shower is matched to the jets from the hard quantum chromodynamics (QCD)
 matrix element via the MLM prescription~\cite{Hoche:2006ph}
with a transverse momentum (\pt) threshold of $20\GeV$. The CMS detector response is simulated using \GEANTfour~\cite{Agostinelli:2002hh}.

Independent \ttbar samples are also generated at both \sqrtSeven and \sqrtEight with \POWHEG v2 r2819~\cite{POWHEG1:Nason:2004rx,POWHEG2:Frixione:2007vw,POWHEG3:Alioli:2010xd}.
At 8\TeV, additional samples are generated with both \MCATNLO v3.41~\cite{Frixione:2002ik}
and \POWHEG v1.0 r1380~\cite{POWHEG1:Nason:2004rx,POWHEG2:Frixione:2007vw,POWHEG3:Alioli:2010xd}.
All of the \POWHEG samples are interfaced with both \PYTHIA and \HERWIG v6.520~\cite{Herwig6}, whereas the \MCATNLO generator
is interfaced with \HERWIG for parton showering.  These samples, which are all generated to next-to-leading order accuracy,
are used for comparison with the final results.

The most significant backgrounds to \ttbar production are events in which a $\PW$ boson is produced in association
with additional jets.
Other backgrounds include single top quark production, $\PZ$ boson production in association with
multiple jets, and QCD multijet events where hadronic activity is misidentified as a lepton.
The simulation of background from $\PW$ and $\PZ$ boson production in association with jets is also performed using
the combination of \MADGRAPH and \PYTHIA, with a \pt\ matching threshold of 10\GeV in this case.
These samples are referred to as $\PW$+jets and $\PZ$+jets, respectively.
Single top quark production via $t$- and $s$-channel $\PW$ boson exchange~\cite{singletopPOWHEG:stchannel}
and with an associated on-shell $\PW$ boson~\cite{singletopPOWHEG:tWchannel} are generated using \POWHEG.
The QCD multijet processes are simulated using \PYTHIA.
The event yields of the background processes are normalized according to their predicted production \xsect\ values.
These are from NNLO calculations for $\PW$+jets and $\PZ$+jets events~\cite{fewz,Wproduction},
next-to-leading order calculations with NNLL corrections for single top quark events~\cite{singletopxsec},
and leading-order calculations for QCD multijet events~\cite{pythia}.

Samples are generated using the {CTEQ6L} PDFs~\cite{Pumplin:2002vw}
for \MADGRAPH samples, the {CT10} PDFs~\cite{ct10} for \POWHEG samples, and the {CTEQ6M}
PDFs~\cite{Pumplin:2002vw} for \MCATNLO.
The \PYTHIA Z2 tune is used to describe the underlying event in both the \MADGRAPH and \POWHEGPYTHIA samples at \sqrtSeven,
whereas the Z2* tune is used for the corresponding samples at \sqrtEight~\cite{Z2UE}.
The underlying event in the \POWHEGHERWIG samples is described by the AUET2 tune~\cite{AUET2UE},
whereas the default tune is used in the \MCATNLOHERWIG sample.

The value of the top quark mass is fixed to $\mtop = 172.5\GeV$ in all samples.
In all cases, \PYTHIA is used for simulating the gluon radiation and fragmentation, following the prescriptions of
Ref.~\cite{bib:io}. Additional simulated hadronic $\Pp\Pp$ interactions (``pileup''), in the same or nearby beam crossings,
are overlaid on each simulated event to
match the high-luminosity conditions in actual data taking.

Previous measurements of differential \ttbar production cross sections at the LHC~\cite{CMSDiffTop7TeV, CMSDiffTop8TeV,ATLASDiffTop7TeV}
showed that several of the \ttbar event generators considered in this analysis predict a harder top quark \pt spectrum
than that observed in data.  An additional simulated \ttbar sample is considered here, where the sample produced with the \MADGRAPH event generator
is reweighted to improve the agreement of the top quark \pt spectrum with data.

\section{Event reconstruction and selection}
\label{sec:selection}

Parallel selection paths for the two lepton types are implemented, resulting
in samples classified as electron+jets and muon+jets.
The trigger for the electron+jets channel during the \sqrtSeven data taking
selects events containing an electron candidate with $\pt>25\GeV$
and at least three reconstructed hadronic jets with $\pt>30\GeV$. In the \sqrtEight data,
at least one electron candidate with $\pt > 27\GeV$ is required, with
no additional requirement for jets.
In the muon+jets channel, at least one isolated muon candidate with $\pt > 24\GeV$ is required at the trigger
level. Each candidate event is required to contain at least one well-measured vertex~\cite{Chatrchyan:2014fea},
located within the $\Pp\Pp$ luminous region in the center of CMS.

Events are reconstructed using a particle-flow (PF) technique~\cite{bib:pf2009,bib:pf2010}, which combines
information from all subdetectors to optimize the reconstruction and identification of
individual long-lived particles.

Electron candidates are selected with a multivariate technique
using calorimetry and tracking information~\cite{Khachatryan:2015hwa}.
Inputs to the discriminant include information about the calorimeter shower shape, track quality,
track-shower matching, and a possible photon conversion veto.
Electron candidates are required to have $\et>30\GeV$ and pseudorapidity in the range $|\eta|<2.5$.
The low-efficiency region $1.44 < |\eta| < 1.57$ between
the barrel and endcap sections of the detector is excluded.
Muon candidates are selected with tight requirements on track and vertex quality, and on
hit multiplicity in the tracker and muon detectors~\cite{Chatrchyan:2012xi}.
These requirements suppress cosmic rays, misidentified muons, and nonprompt muons from decay of hadrons in flight.
Muon candidates are required to have $\pt>26\GeV$ and $|\eta|<2.1$.

For the lepton isolation requirement, a cone of size $\Delta R=\sqrt{\smash[b]{(\Delta\eta)^2 + (\Delta\phi)^2}}$ is
constructed around the lepton direction, where $\Delta\eta$ and $\Delta\phi$ are the differences in
pseudorapidity and azimuthal angle (in radians), respectively,
between the directions of the lepton and another particle.
The \pt values of charged and neutral particles found in this cone are summed,
excluding the lepton itself and correcting for the effects of pileup~\cite{Khachatryan:2015hwa}.
The relative isolation variable $I(\Delta R)$ is defined as the ratio of this
sum to the lepton \pt.
Lepton candidates are selected if they satisfy $I(0.3) < 0.1$ for electrons,
and $I(0.4) < 0.12$ for muons.

Reconstructed particles are clustered into jets using the
anti-$\kt$ algorithm~\cite{bib:antikt} with a distance parameter of 0.5.
The measured \pt of each jet is corrected~\cite{Chatrchyan:2011ds} for known
variations in the jet energy response as a function of the measured
jet $\eta$ and \pt.
The jet energy is also corrected for the extra energy deposition from
pileup interactions~\cite{bib:PUSubtraction,Cacciari:2007fd}.
Jets are required to
pass loose identification
requirements to remove calorimeter noise~\cite{Chatrchyan:2013txa}.
Any such jet whose direction is less than $\Delta R = 0.3$ from the
identified lepton direction is removed.
For the identification of $\PQb$ quark jets (``$\PQb$ tagging''),
a ``combined secondary vertex'' algorithm~\cite{bib:btag2012} is used,
taking into account the reconstructed secondary vertices and track-based lifetime information.
The $\PQb$ tagging threshold is chosen to give an acceptance of 1\%
for light-quark and gluon jets with a tagging efficiency of 65\% for $\PQb$ quark jets.

The final selection requires exactly one high-\pt, isolated electron or
muon.
Events are vetoed if they contain an additional lepton candidate
satisfying either of the following criteria: an electron with $\pt>20\GeV$,
$|\eta|<2.5$, and $I(0.3) < 0.15$; or a
muon, with looser requirements on hit multiplicity, and with $\pt>10\GeV$, $|\eta|<2.5$, and $I(0.4) <
0.2$.  The event must have at least four jets with $\pt >30\GeV$, of which at
least two are tagged as containing $\PQb$ hadrons.

After the final selection, 26~290 data events are found at \sqrtSeven, and 153~223 at \sqrtEight.
The \ttbar contribution to these event samples, as estimated from simulation, is about 92\%.
The fraction of true signal events in the samples is 78\%.
Misidentified all-hadronic or dileptonic \ttbar events, and events containing tau leptons among the \ttbar decay products,
comprise 14\% of the samples.
The remaining events are approximately 4\% single top quark events, 2\% $\PW$/$\PZ$+jets events,
and 2\% QCD multijet events.  The efficiency for signal events to satisfy the final selection
criteria is about 8\%, as determined from simulation.

\section{\xSect\ measurements}
\label{sec:cross_section}

We study the normalized \ttbar differential production cross section
as a function of four kinematic event variables: \met, \HT, \st,
and \wpt.

The variable \met\ is the magnitude of the missing transverse momentum vector \ptvecmiss,
which is defined as the projection on the plane
perpendicular to the beams of the negative vector sum of the momenta of all PF candidates in the
event:
\begin{equation*}
\met = \left[ \left(-\sum_i{p_{x}^i}\right)^2 + \left(-\sum_i{p_{y}^i}\right)^2 \right]^{\frac{1}{2}},
\end{equation*}
where $p_x^i$ and $p_y^i$ are the $x$ and $y$ momentum components of the \textit{i}th candidate,
and the sums extend over all PF candidates.  The measured \met\ is corrected for pileup and
nonuniformities in response as a function of $\phi$~\cite{Khachatryan:2014gga}.

The variable \HT is defined as the scalar sum of the transverse momenta of all jets in the event,
\[\HT = \sum_{\text{all jets}}\pt^{\mathrm{jet}} ,\]
where the sum extends over all jets having  $\pt>20\gev$ and $\left|\eta\right|<2.5$.

The variable \st\ is the scalar sum of \HT, \met,\ and the \pt of the identified lepton,
\[\st = \HT + \met + p_\mathrm{T}^\mathrm{lepton}.\]

Finally, \wpt\ is the magnitude of the transverse momentum of the leptonically decaying $\PW$ boson,
which is derived from the momentum of the isolated lepton and  \ptvecmiss\
\[\wpt = \sqrt{\left(p_x^{\mathrm{lepton}} + p_x^{\mathrm{miss}}\right)^2 +
\left(p_y^{\mathrm{lepton}} + p_y^{\mathrm{miss}}\right)^2} ,\]
where $p_{x}^{\mathrm{lepton}}$ and $p_{y}^{\mathrm{lepton}}$ are the transverse components of
$\vec{p}^{\mathrm{lepton}}$, and $p_x^{\mathrm{miss}}$ and $p_y^{\mathrm{miss}}$ are the transverse components
of \ptvecmiss.

\begin{figure*}[hbtp]
\begin{center}
\includegraphics[width=\cmsFigWidth]{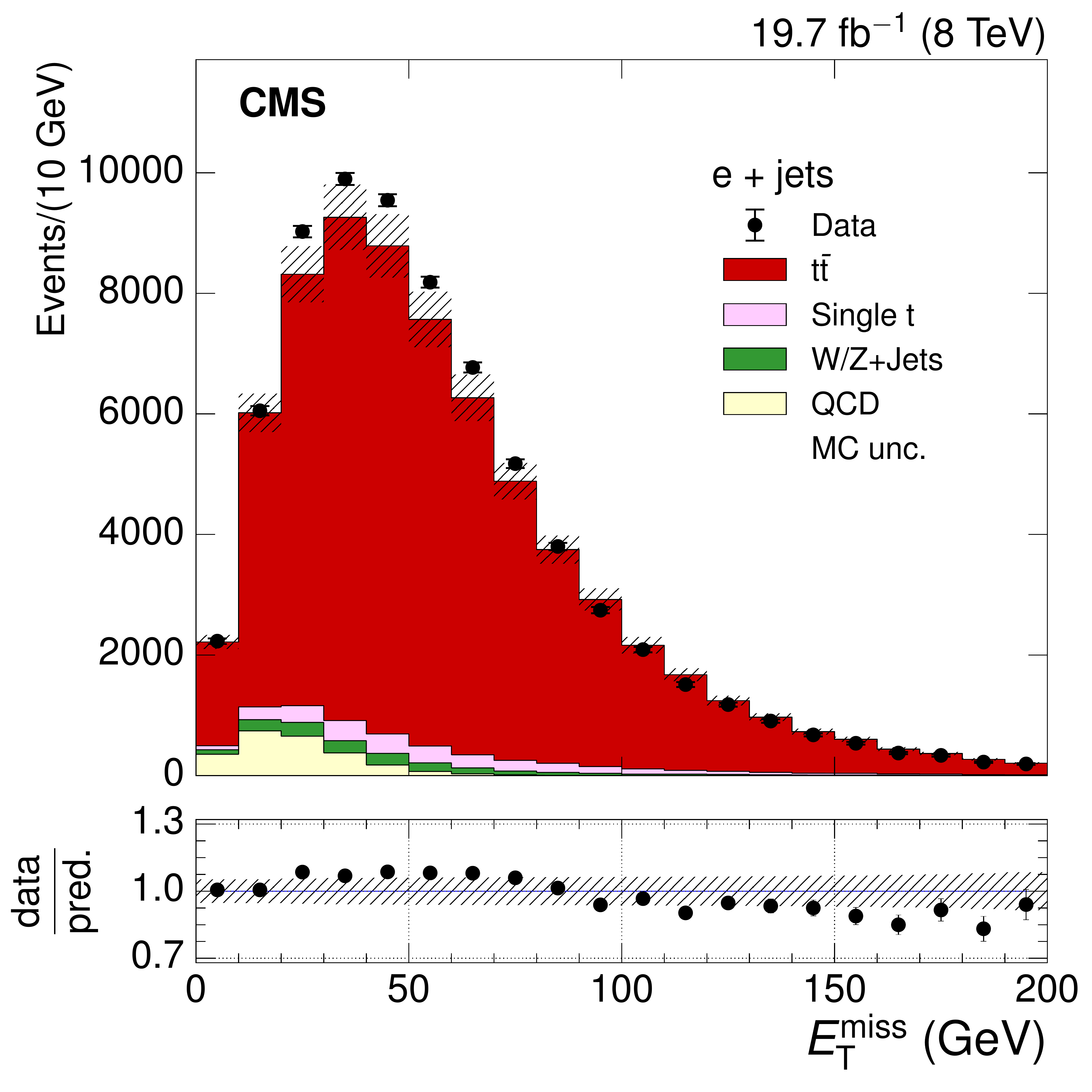}\hfill
\includegraphics[width=\cmsFigWidth]{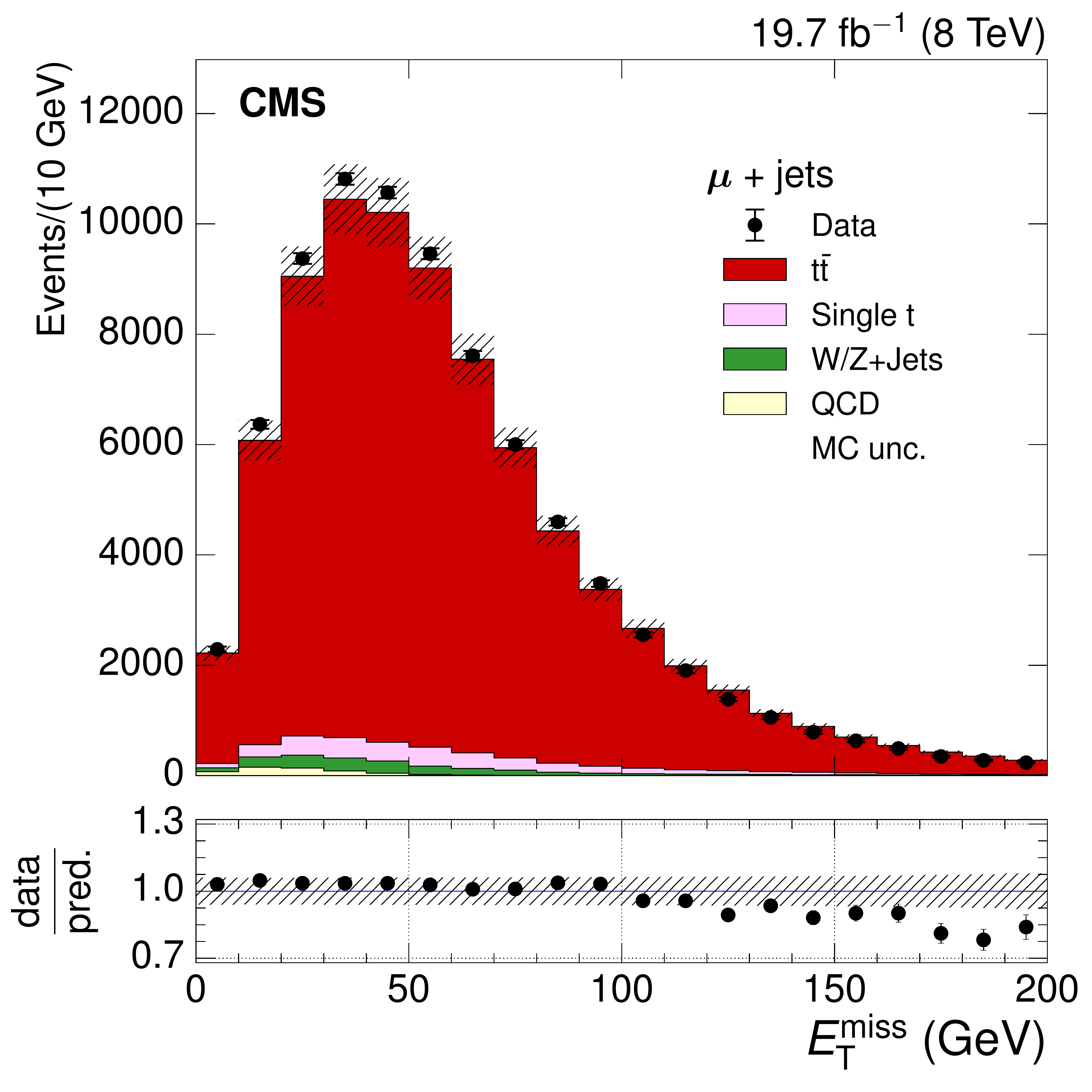}\\
\includegraphics[width=\cmsFigWidth]{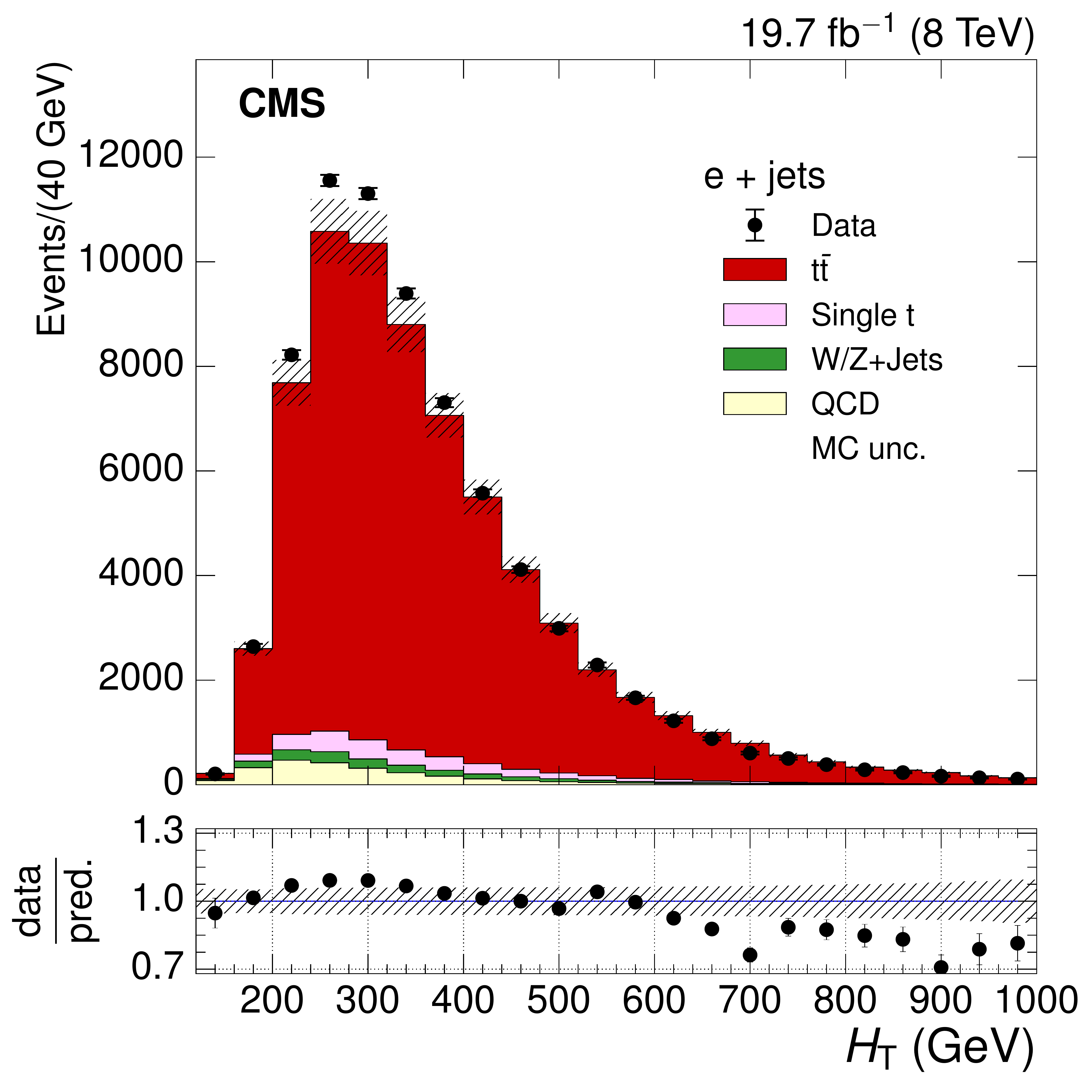}\hfill
\includegraphics[width=\cmsFigWidth]{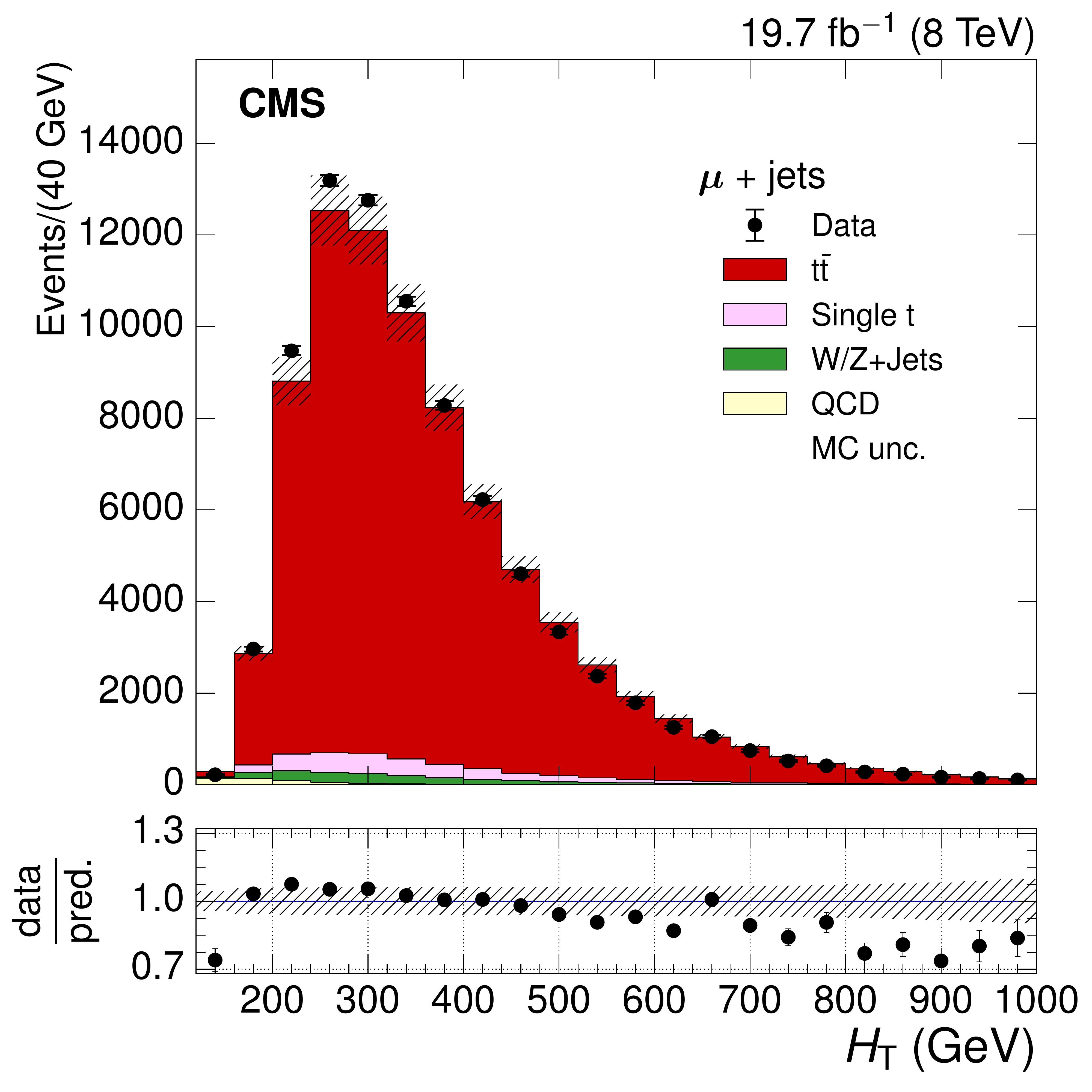}
\caption{\label{fig:control_MET_HT}The observed distributions of
\met\ (top) and \HT (bottom) in the $\sqrt{s}=8\TeV$
electron+jets (left) and muon+jets (right) data samples,
compared to predictions from simulation.
The points are the data histograms, with the vertical bars showing the statistical uncertainty,
and the predictions from the simulation are the solid histograms.
The shaded region shows the uncertainty in the values from simulation.
These include contributions from the statistical uncertainty and the uncertainty in the \ttbar cross section.
The lower plots show the ratio of the number of events from data
and the prediction from the MC simulation.}
\end{center}
\end{figure*}

\begin{figure*}[hbtp]
\begin{center}
\includegraphics[width=\cmsFigWidth]{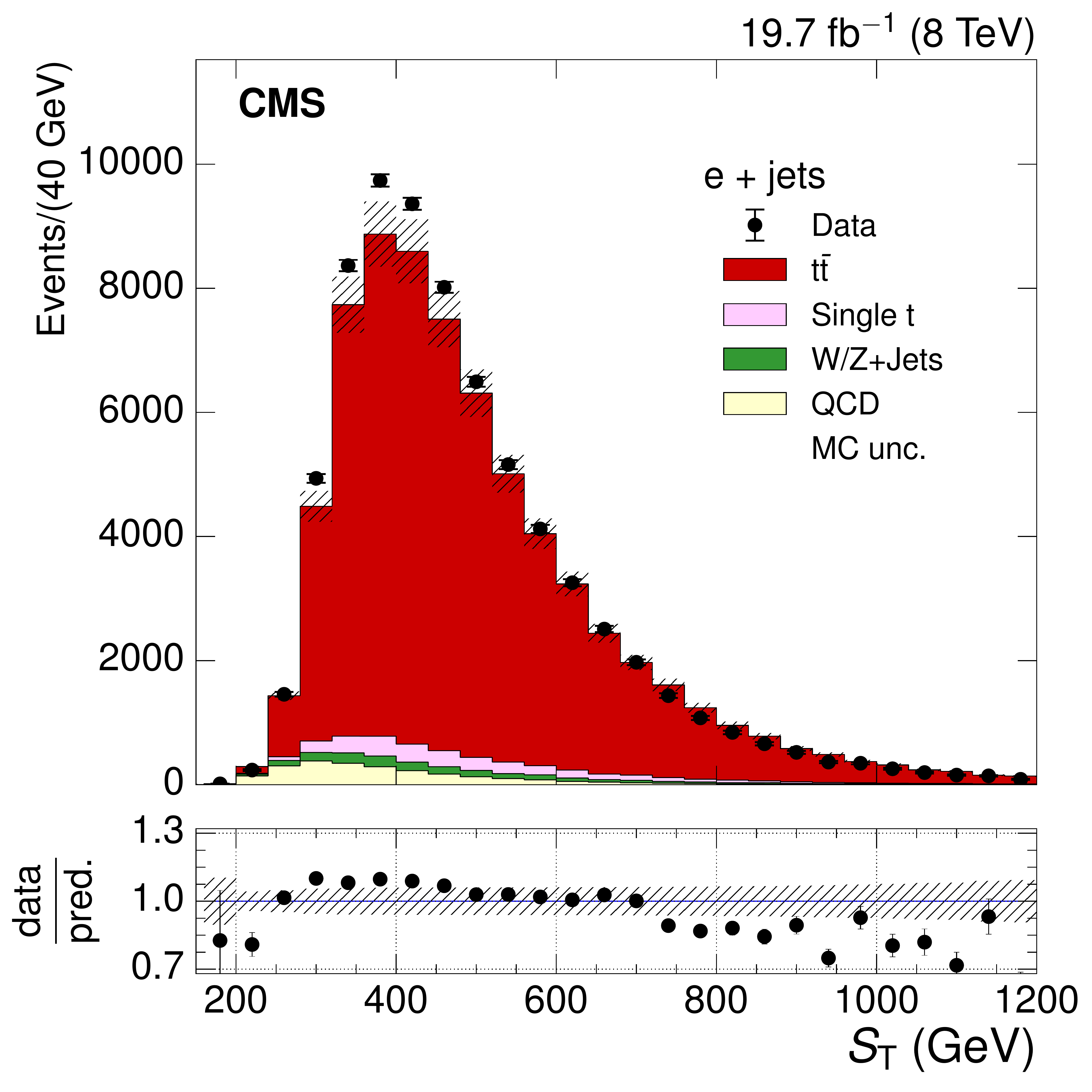}\hfill
\includegraphics[width=\cmsFigWidth]{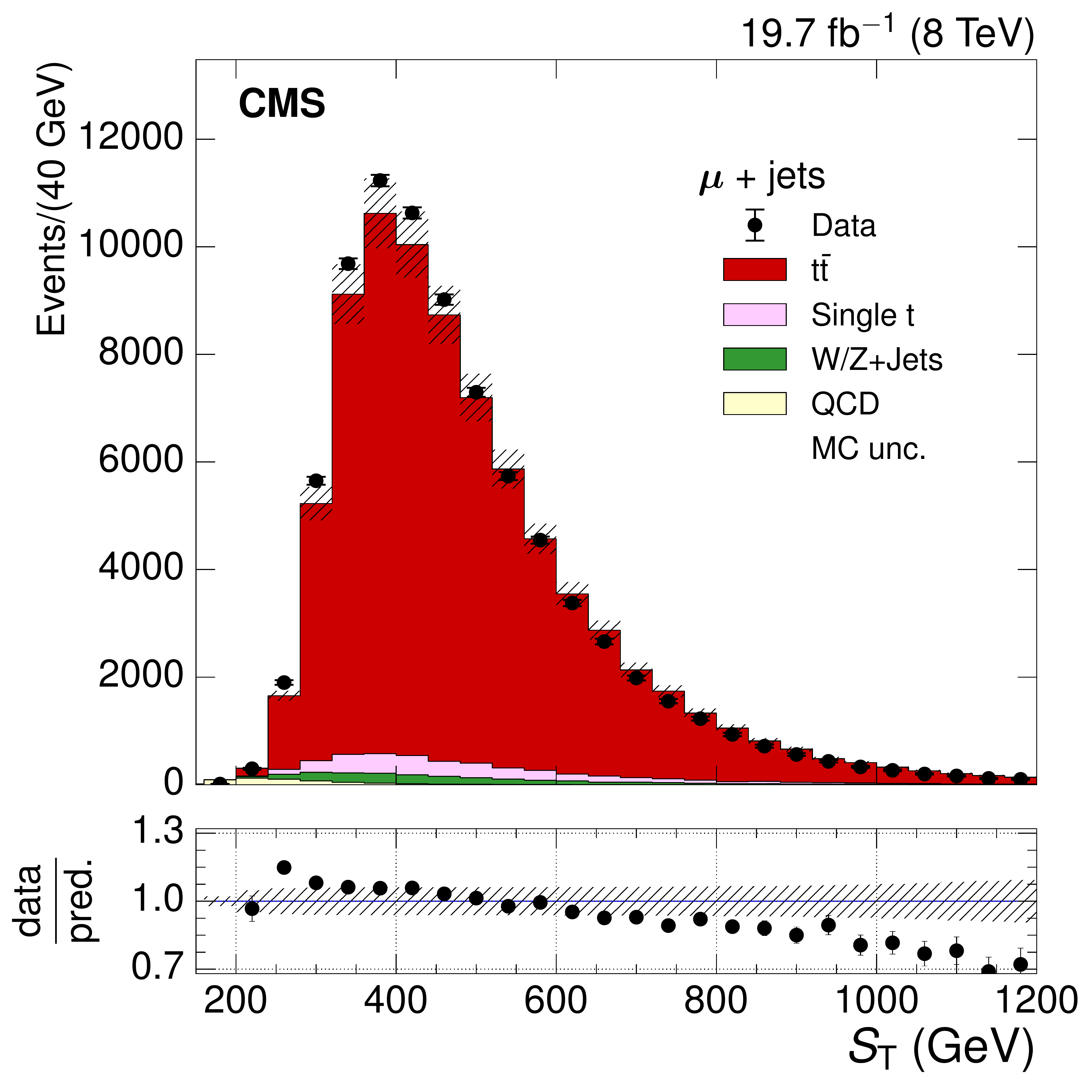}\\
\includegraphics[width=\cmsFigWidth]{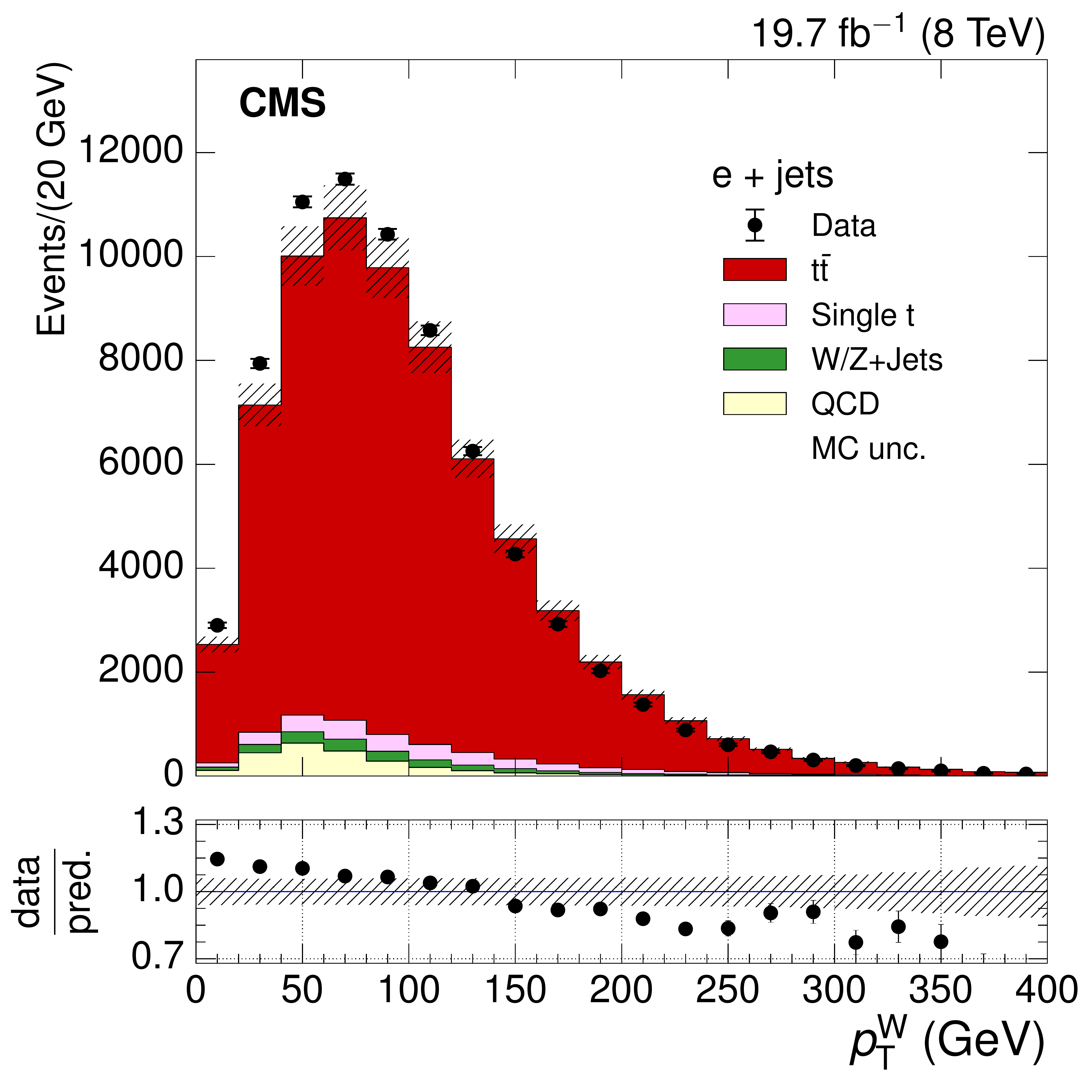}\hfill
\includegraphics[width=\cmsFigWidth]{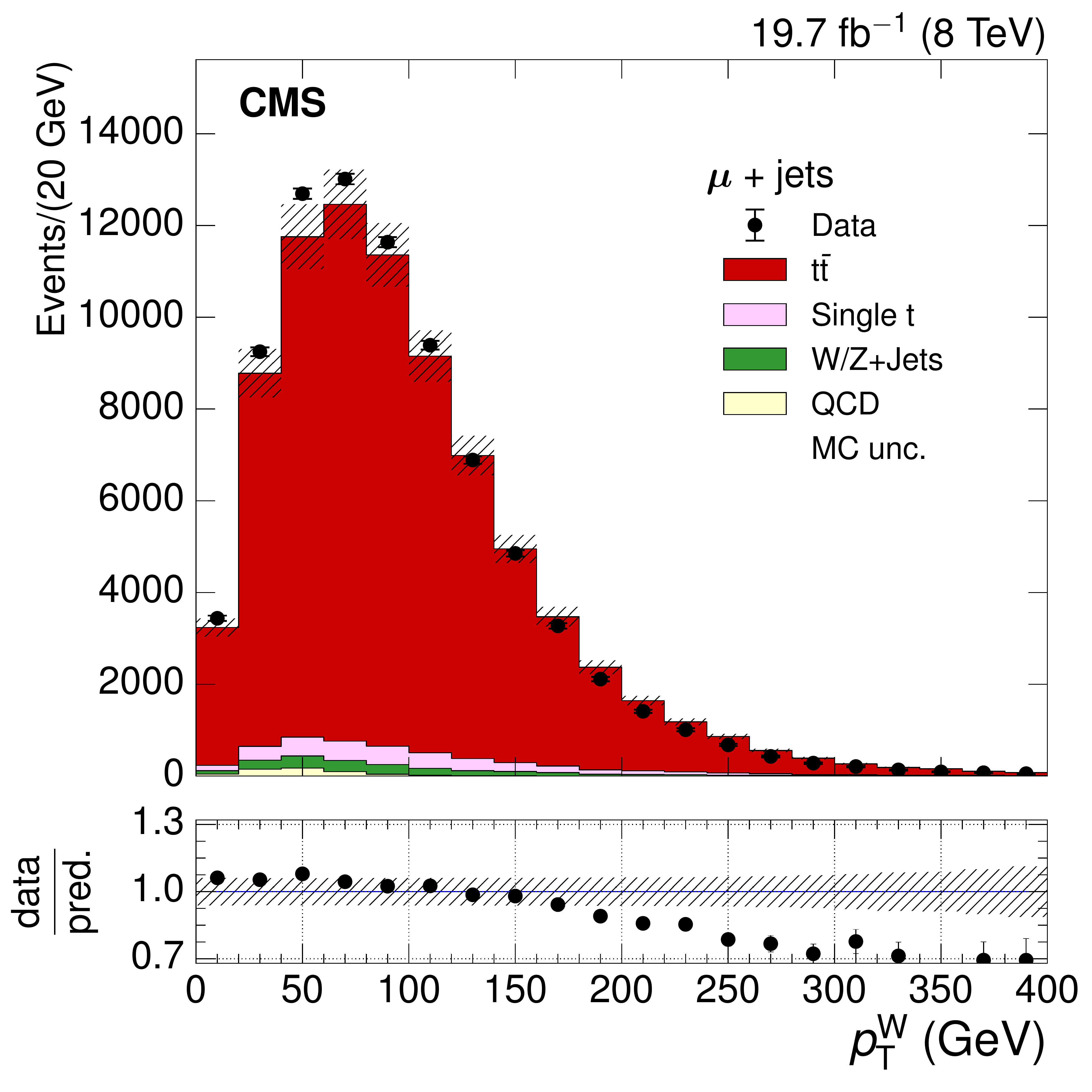}
\caption{\label{fig:control_ST_WPT}The observed distributions of
\st\ (top) and \wpt (bottom) in the $\sqrt{s}=8\TeV$
electron+jets (left) and muon+jets (right) data samples,
compared to predictions from simulation.
The points are the data histograms, with the vertical bars showing the statistical uncertainty,
and the predictions from the simulation are the solid histograms.
The shaded region shows the uncertainty in the values from simulation.
These include contributions from the statistical uncertainty and the uncertainty in the \ttbar cross section.
The lower plots show the ratio of the number of events from data
and the prediction from the MC simulation.}
\end{center}
\end{figure*}

Figures~\ref{fig:control_MET_HT}~and~\ref{fig:control_ST_WPT} show the observed distributions of
\met, \HT, \ST, and \wpt, in the \sqrtEight data samples,
compared to the sum of the corresponding signal and background distributions from simulation.

For simulated \tt\ signal events, these four kinematic variables are also calculated
using the momenta of particles in the event, before the simulation of the detector response.
We refer to the quantities calculated in this way as the generated variables.
The generated value of \met\ is the magnitude of the vector sum of the \pt\ of all neutrinos in the event.
The long-lived particles in the event are clustered into jets in the same way as the reconstructed
particles.
The generated value of \HT\ is the sum of the magnitudes of the \pt\ of these jets with $\pt>20\gev$ and $\left|\eta\right|<2.5$.
The generated values of \st\ and \wpt\ are calculated in the same way as the corresponding reconstructed variables,
using the \ptvec\ of the charged lepton from the leptonic decay of a $\PW$ boson coming from $\PQt \to \PQb \PW$ decay.

The choice of bin widths for this measurement is optimized separately for each kinematic event variable
to minimize the migration between bins. This optimization is based on three criteria:
(i) of the simulated signal events for which the value of the generated variable falls in the bin,
at least 50\% are required to have the reconstructed variable in the same bin (this is sensitive to migration of
events out of the bin);
(ii) of the simulated signal events for which the value of the reconstructed variable falls in the bin,
at least 50\% are required to have the generated variable in the same bin (this is sensitive to migration of
events into the bin);
(iii) the number of reconstructed
simulation events in a bin is required to be more than 100.
These criteria ensure that bin-to-bin migrations are kept small, while allowing a differential
cross section measurement with reasonable granularity.

The number of \ttbar events in each bin of each kinematic event variable, and in each channel, is obtained by subtracting the expected
contributions of background processes from data.  The contributions of single top quark, and $\PW$ or $\PZ$ boson plus jet events are estimated
from simulation.

In the case of the QCD multijet background, the contribution is estimated from data using a control region where the
selection criteria are modified to enrich the contribution of QCD multijet events.
In the electron+jets channel, the control region is obtained by inverting the photon conversion veto on the electron.
In addition to this, the number of $\PQb$-tagged jets is required to be exactly zero.
The small contamination of \tt, single top, $\PW$+jets, and $\PZ$+jets events in this control region, as estimated from simulation,
is subtracted from the data.
Then, the ratio of simulated QCD multijet events in the control region and the signal region is used to scale
the normalization of the data-driven QCD multijet estimate from the control region to the signal region in the data.
The control region in the muon+jets channel is obtained by inverting the isolation criterion on the muon in the selected events,
and by requiring exactly zero $\PQb$-tagged jets.  The jet selection criterion is also modified,
requiring at least three jets.
The same procedure is then followed to estimate the contribution of QCD mulitjet events in the muon+jets signal region.

The number of \ttbar events from data in each bin is then corrected
for the small fractions of dileptonic, all-hadronic, and tau \ttbar events in the
final sample, as determined from simulation,
and for experimental effects, such as detector resolution, acceptance, and
efficiency. This correction is performed by constructing a
response matrix that maps the generated values to the reconstructed values for the
four kinematic variables in the simulated \ttbar signal events.  The response
matrix is constructed using the \MADGRAPH \ttbar sample.
This matrix is then inverted, using regularized singular-value
decomposition \cite{SVD_Unfolding} in the {\sc RooUnfold} \cite{Adye:2011gm}
software framework. Since we impose no
requirements on the generated events, the procedure corrects to the
full signal phase space.

The fully-corrected numbers of \ttbar events in the electron+jets and
muon+jets channels yield consistent results.  These are then added and used to calculate the
normalized \ttbar differential production cross section with respect to each kinematic event variable, $X$, using:
\begin{equation}
\frac{1}{\sigma}\frac{\rd \sigma_j}{\rd X}=\frac{1}{N}\frac{x_{j}}{\Delta_{j}^{X}}\;,
\end{equation}
where $x_{j}$ represents the number of unfolded signal events in bin $j$,
$\Delta_{j}^{X}$ is the width of bin $j$; $\sigma$ is the total \ttbar production cross section,
and $N=\sum_{i}{x_{i}}$ is the total number of unfolded signal events.

\section{Systematic uncertainties}
\label{sec:systematic_errors}

The systematic uncertainties in the experimental and theoretical input quantities
are evaluated and propagated to the final results, taking correlations
into account. Since the final result is normalized to the total number of events,
the effect of uncertainties that are correlated across all bins
is negligible.
As such, only uncertainties that affect the shape of the measured distributions are significant.

The uncertainty coming from the choice of
renormalization and factorization scales in the physics modeling
of \ttbar events is determined by producing two additional simulated event samples. These samples are generated with
both scales simultaneously varied by a factor of two up or down from their default values
equal to the $Q$ of the hard process in the event;
$Q$ is defined via $Q^2 = \mtop^{2}+\sum{p_{\mathrm{T}}^{2}}$,
where the sum is over all additional final-state partons in the matrix element.
The effect of varying the renormalization and factorization scales
in the $\PW$+jets and $\PZ$+jets samples is also considered to determine the uncertainty in the shape of this background.
The uncertainty arising from the choice of parton shower matching threshold in the event generation
is determined in a similar fashion, using additional samples in which the threshold is varied up or down.
Uncertainties from the
modeling of the hadronization are evaluated by comparing \POWHEG v1 simulated samples
with two different hadron shower generators (\PYTHIA and \HERWIG).
The uncertainty owing to the choice of the PDF is determined by reweighting the
simulated events and repeating the analysis using the 44 {CTEQ6L} PDF error sets~\cite{Pumplin:2002vw}.
The maximum variation is taken as the uncertainty.
Simulated samples with the top quark mass varied by $\pm1\GeV$, which corresponds to the precision of the measured
top quark mass~\cite{PDG}, are generated to evaluate the effect of the uncertainty in this parameter.
The effect of reweighting the top quark \pt spectrum in simulation, as described in Section~\ref{sec:mc_modelling},
is found to have a negligible effect for low values of the kinematic event variables, and
increases to 3--7\% for the highest values.

Other uncertainties are associated with imperfect understanding
of efficiencies, resolutions, and scales describing the detector response.
The uncertainty arising from each source is estimated, and the analysis repeated with
each corresponding parameter varied within its uncertainty.

The efficiencies and associated uncertainties for triggering and lepton
identification are determined from data by a tag-and-probe
method~\cite{Khachatryan:2010xn}.
The probabilities for identifying and misidentifying $\PQb$ jets in the simulation
are compared to those measured in data, and the resulting correction factors and
their uncertainties are determined as a function of jet energy and quark
flavor. The uncertainties in the correction factors are typically 2\%.

The uncertainty in the jet energy scale (JES) is determined as a function of the jet \pt
and $\eta$~\cite{Chatrchyan:2011ds}, and an uncertainty of $10\%$
is included in the jet energy resolution (JER)~\cite{Chatrchyan:2011ds}.  The effect of this limited knowledge of the
JES and JER is determined by varying the JES and JER in the simulated samples within their uncertainties.
The uncertainty in the JES and JER, as well as uncertainties in the
electron, photon, tau, and muon energy scale, are propagated into the calculation of \met.
The uncertainty in the electron and photon energy scale is $0.6\%$ in
the barrel, and $1.5\%$ in the endcap \cite{Khachatryan:2015hwa}. The uncertainty in the tau lepton
energy scale is estimated to be $\pm3\%$ \cite{CMS:Higgstaupaper},
while the effect of the uncertainty in the muon momentum
measurement is found to be negligible.
A 10\% uncertainty is assigned to the estimate of the nonclustered
energy used in the calculation of \met~\cite{Khachatryan:2014gga}.

The effect of the uncertainty in the level of pileup is estimated by varying the inelastic $\Pp\Pp$ cross section
used in the simulation by $\pm 5\%$~\cite{InelasticPP}.

The uncertainty in the normalization of the background is determined by varying the normalization
of the single top, $\PW$+jets, and $\PZ$+jets processes by $\pm30\%$, and the QCD multijet processes by $\pm100\%$.
The uncertainty in the shape of the QCD multijet distribution in the electron channel is
estimated by using an alternative control region in data to determine the contribution of QCD multijet events.
This uncertainty is found to have a negligible effect.

The dominant systematic effects are caused by the
uncertainties in the modeling of the hadronization and the \ttbar signal.
For illustrative purposes, typical systematic uncertainties in the \sqrtEight results coming from each of
the sources described above are presented in
Table~\ref{tab:typical_systematics_8TeV_combined}. The values shown for each kinematic event variable are the median
uncertainties over all of the bins for that variable.

\begin{table}[htbp]
\renewcommand{\arraystretch}{1.1}
\centering
\topcaption{Typical relative systematic uncertainties in percent (median values)
in the normalized \ttbar differential cross
section measurement as a function of the four kinematic event variables
at a center-of-mass energy of 8\TeV (combination of electron and muon channels).
Typical values of the total systematic uncertainty are also shown.}
\label{tab:typical_systematics_8TeV_combined}
\begin{scotch}{c....}
\multirow{2}{*}{Uncertainty source} & \multicolumn{4}{c}{Relative (\%) } \\
 & \multicolumn{1}{c}{\met} & \multicolumn{1}{c}{\HT} & \multicolumn{1}{c}{\st} & \multicolumn{1}{c}{\wpt} \\
\hline
\begin{tabular}{c}Fact./Renorm. scales\\and matching threshold\end{tabular}& 7.6 & 4.0 & 2.6 & 3.3 \\
Hadronization & 4.3 & 5.0 & 8.5 & 3.0 \\
PDF & 0.5 & 0.6 & 0.6 & 0.4 \\
Top quark mass & 0.4 & 0.7 & 0.8 & 0.3 \\
Top quark $p_\mathrm{T}$ reweighting & 1.4 & 0.9 & 0.6 & 0.6 \\
\begin{tabular}{c}Lepton trigger efficiency\\\& selection\end{tabular}& {<}0.1 & {<}0.1 & {<}0.1 & {<}0.1 \\
$\PQb$ tagging & 0.3 & 0.1 & 0.3 & {<}0.1 \\
Jet energy scale & 0.3 & 0.2 & 0.3 & {<}0.1 \\
Jet energy resolution & {<}0.1 & {<}0.1 & {<}0.1 & {<}0.1 \\
\met & 0.2 & \multicolumn{1}{c}{\NA} & {<}0.1 & 0.1 \\
Pileup & 0.4 & {<}0.1 & 0.1 & 0.2 \\
Background Normalization & 2.6 & 1.0 & 2.1 & 1.4 \\
QCD shape & 0.4 & 0.2 & 0.5 & 0.4 \\
\hline
Total & 9.9 & 8.6 & 9.5 & 4.4 \\
\end{scotch}
\end{table}

\section{Results}
\label{sec:comb_results}

\begin{figure*}[!htb]
  \begin{center}
        \includegraphics[width=\cmsFigWidth]{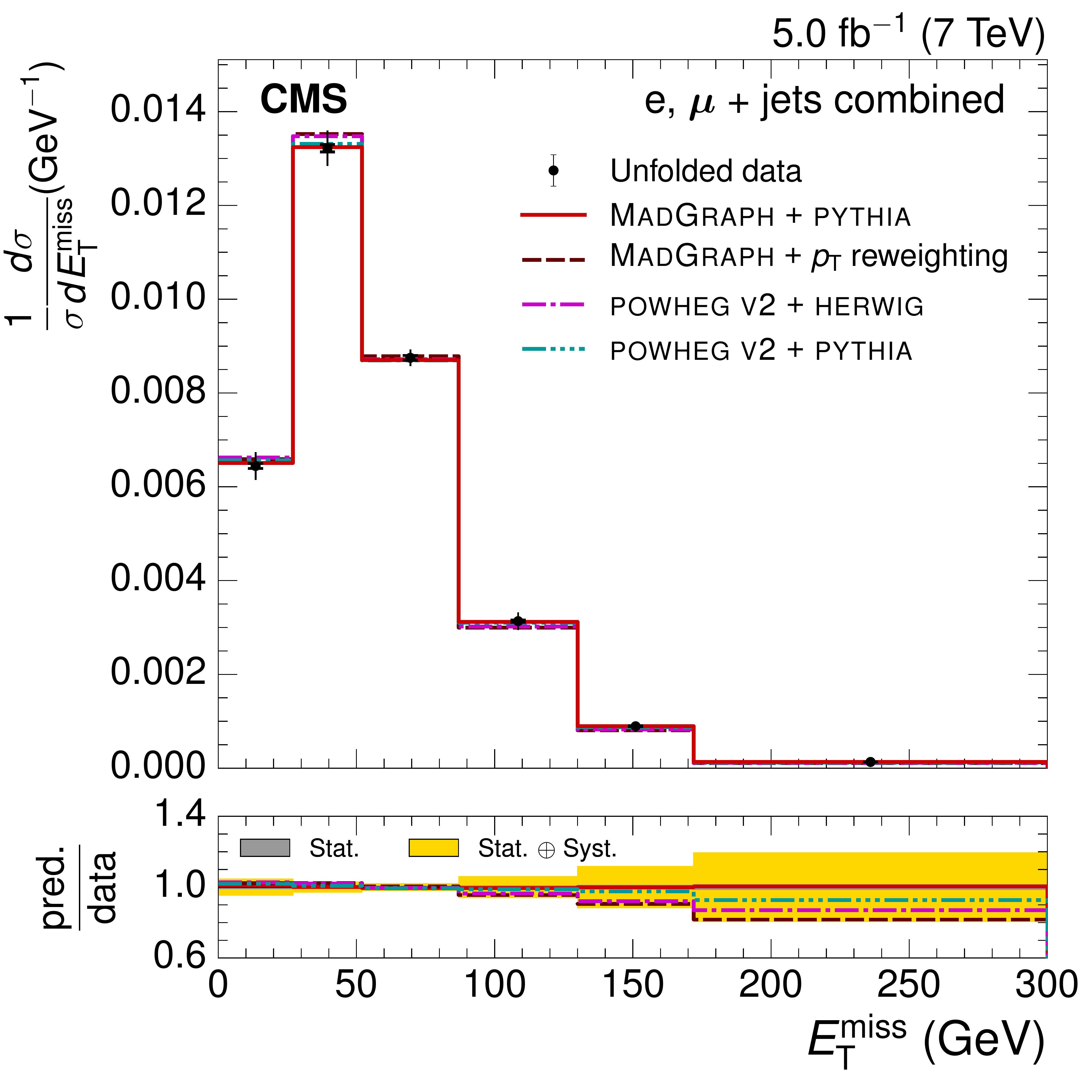} \hfill
        \includegraphics[width=\cmsFigWidth]{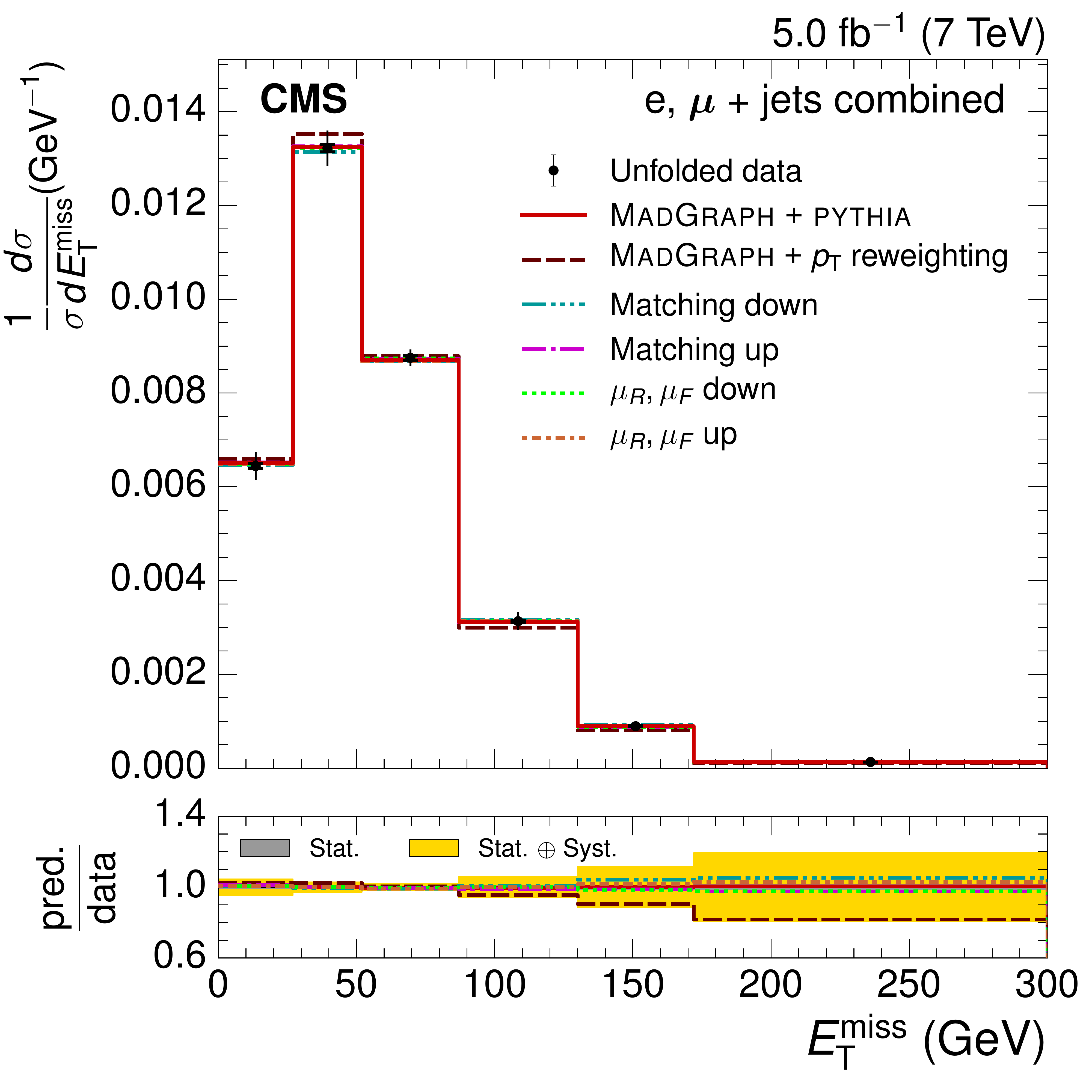} \\
        \includegraphics[width=\cmsFigWidth]{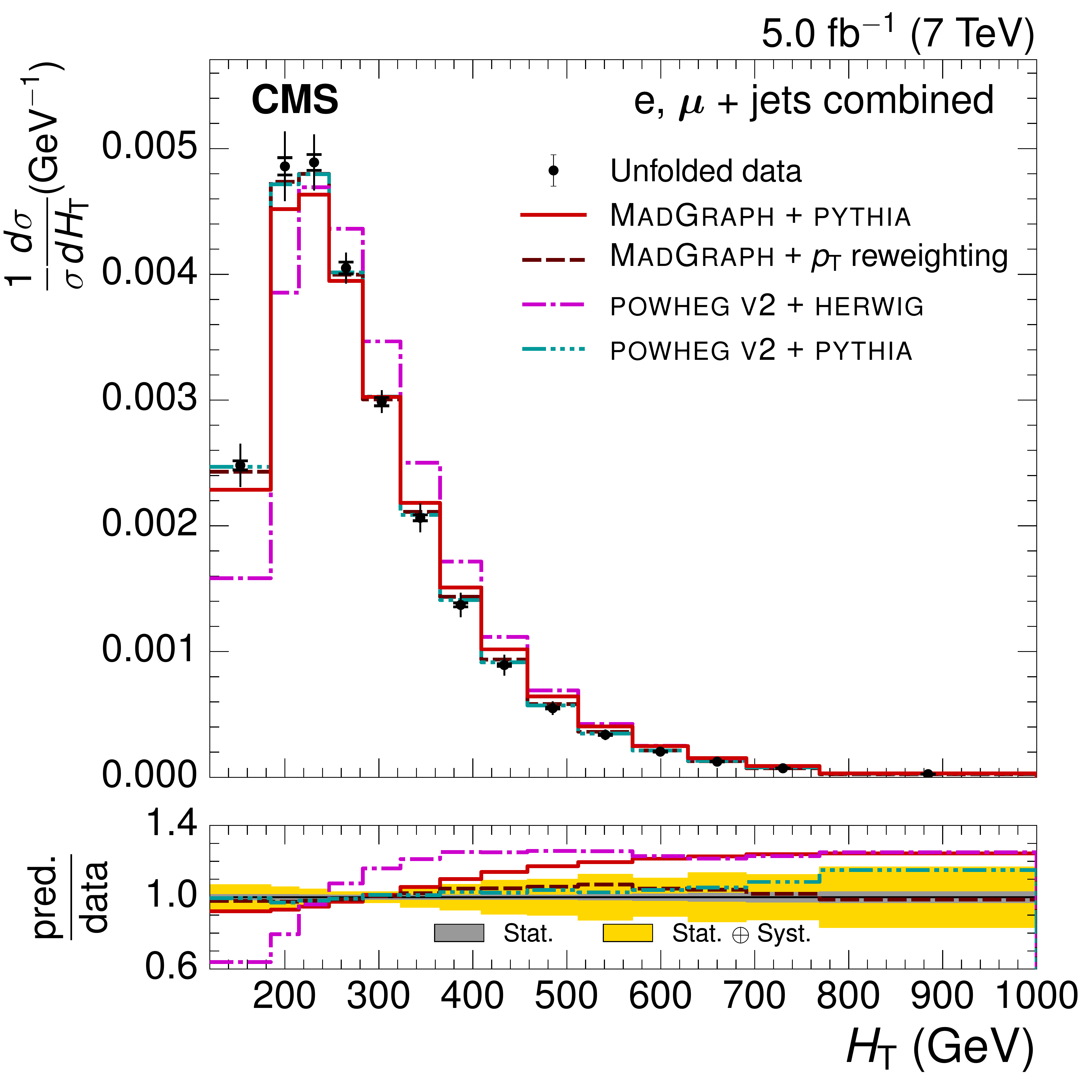} \hfill
        \includegraphics[width=\cmsFigWidth]{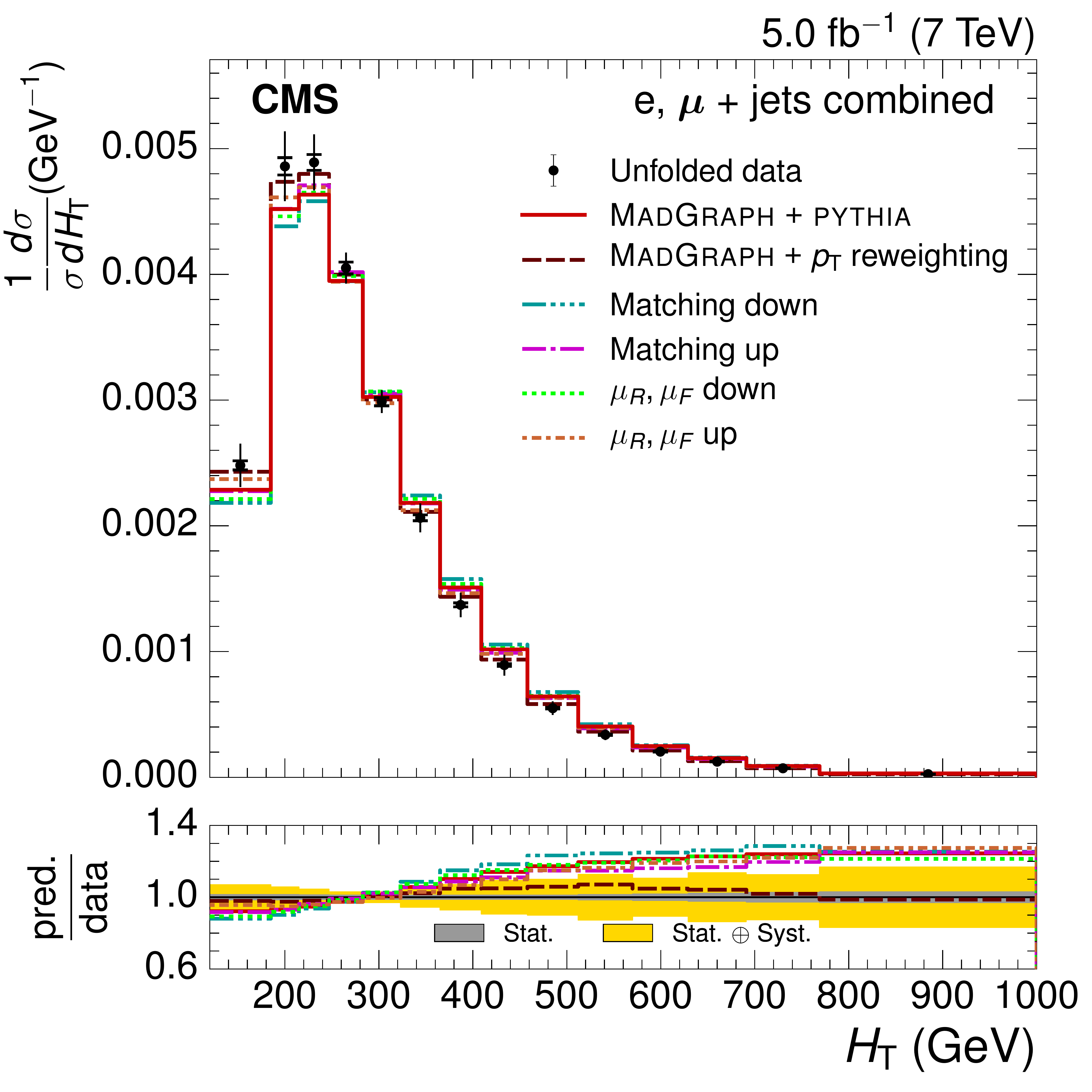}
    \caption{Normalized \met (top) and \HT (bottom) differential \ttbar cross sections from the combined
    electron and muon data at $\sqrt{s}=7\TeV$.
    The vertical bars on the data points represent the statistical and systematic uncertainties added in quadrature.
    The inner section of the vertical bars, denoted by the tick marks, show the statistical uncertainty.
    Left: comparison with different
    simulation event generators: \MADGRAPHPYTHIA (both the default and after reweighting the top quark \pt spectrum),
    \POWHEGVTWOHERWIG, and \POWHEGVTWOPYTHIA. Right: comparison with
    predictions from the \MADGRAPHPYTHIA event generator found by varying the matching threshold and renormalization
    scales ($\mu_{\mathrm{R}}$, $\mu_{\mathrm{F}}$) up and down by a factor of two.  The lower plots show the ratio of the
    predictions to the data, with the statistical and total uncertainties in the ratios indicated by the two shaded bands.}
    \label{fig:MET_HT_result_combined_7TeV}
  \end{center}
\end{figure*}

\begin{figure*}[!htb]
  \begin{center}
        \includegraphics[width=\cmsFigWidth]{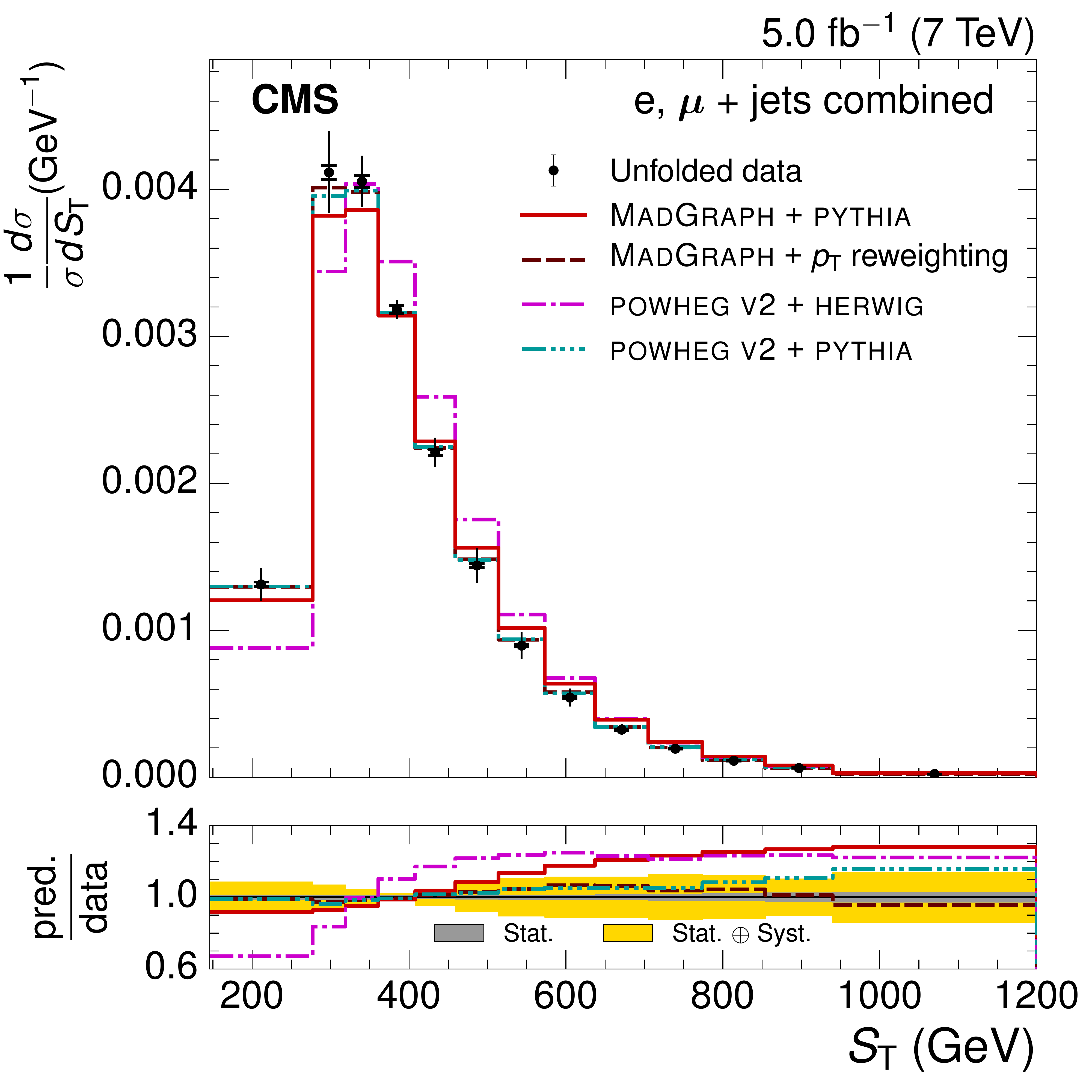}\hfill
        \includegraphics[width=\cmsFigWidth]{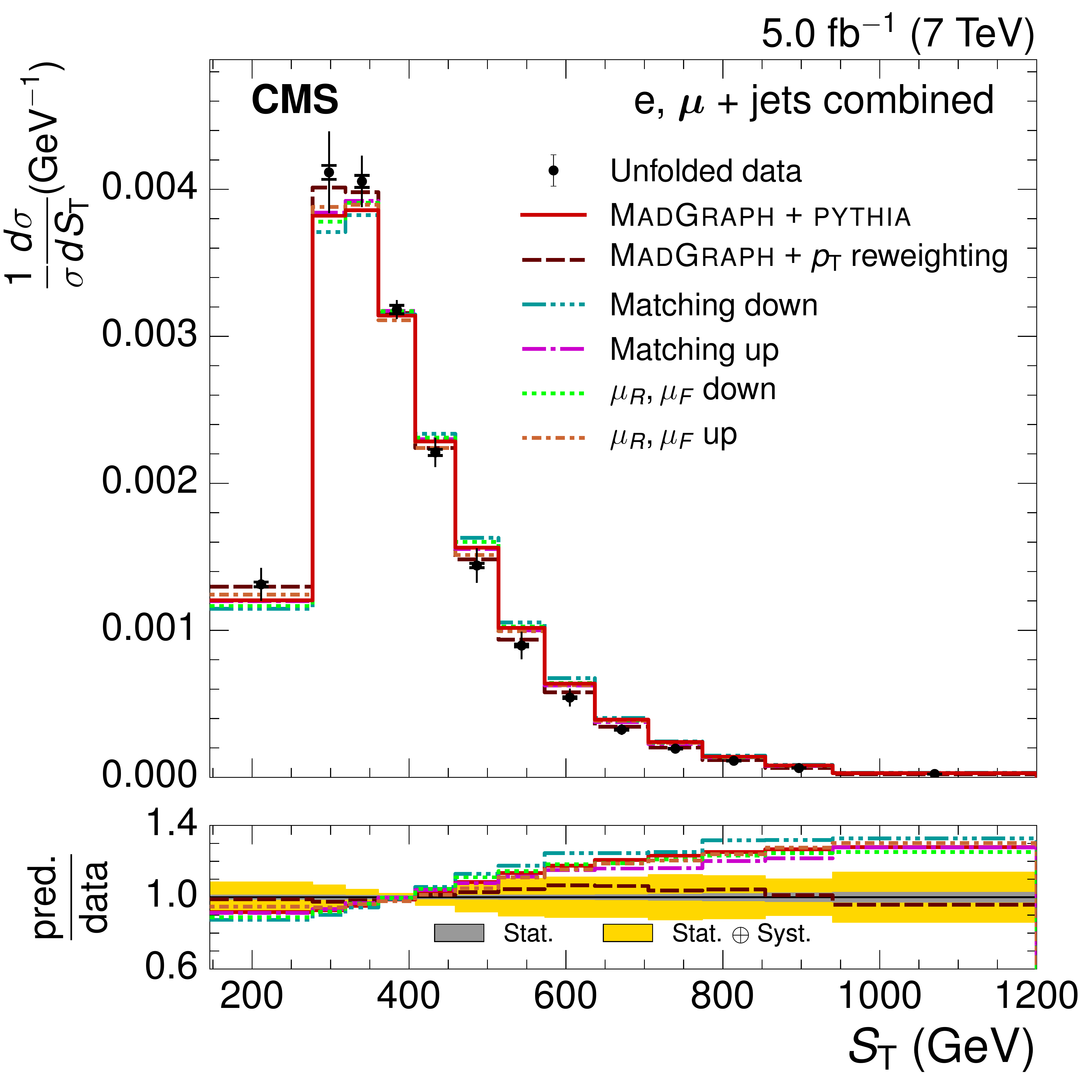} \\
        \includegraphics[width=\cmsFigWidth]{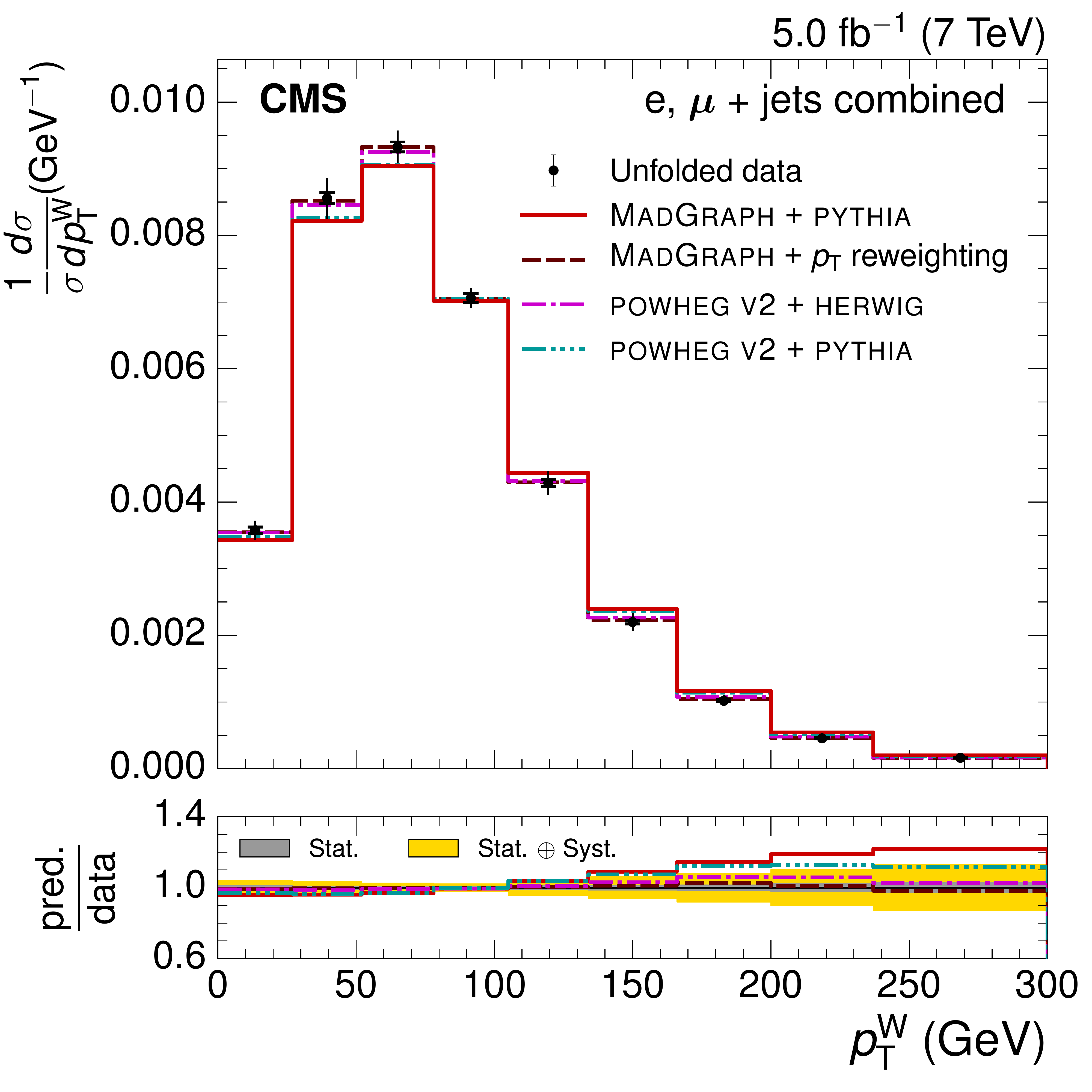}\hfill
        \includegraphics[width=\cmsFigWidth]{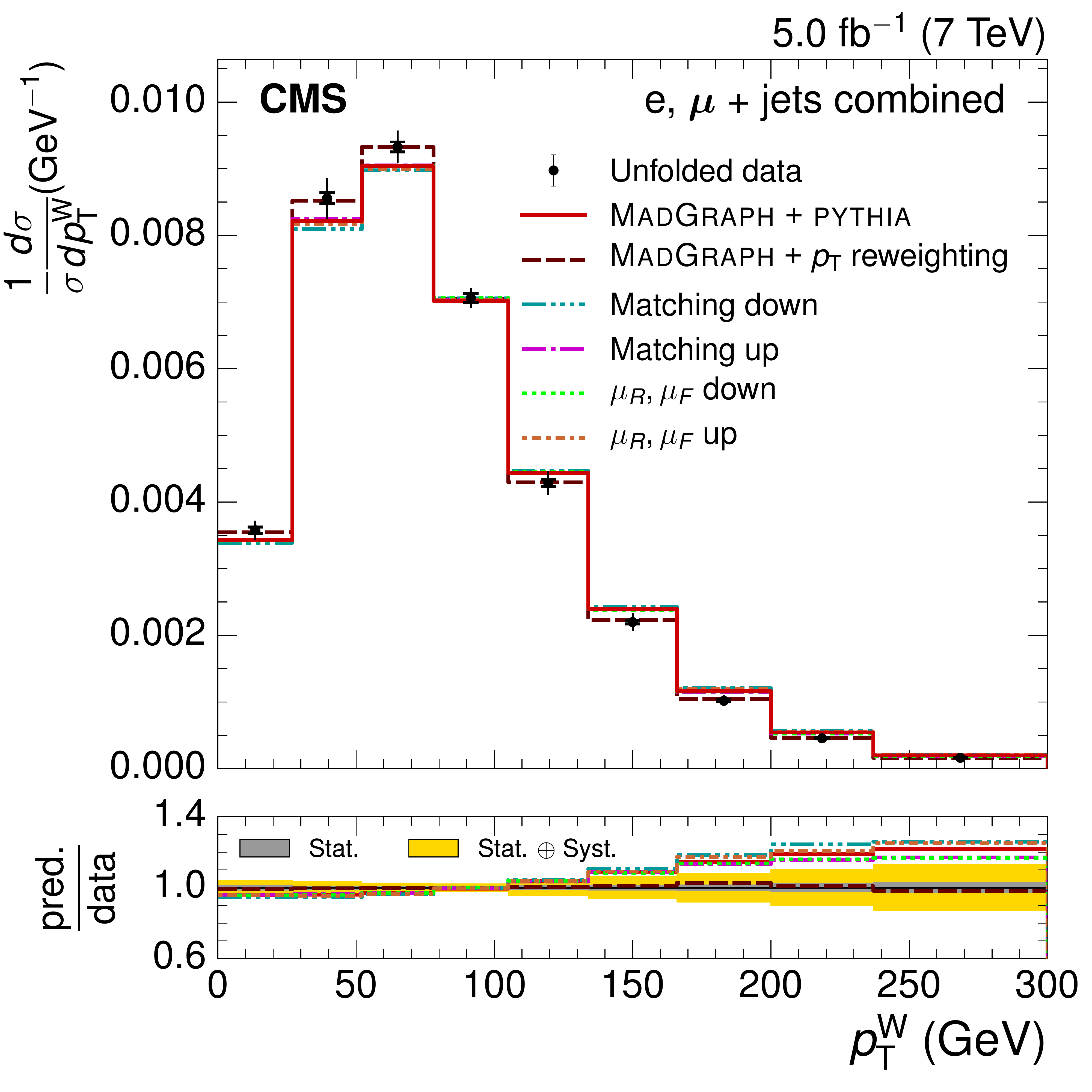}
    \caption{Normalized \st (top) and \wpt (bottom) differential \ttbar cross sections from combined
    electron and muon data at $\sqrt{s}=7\TeV$.
   The vertical bars on the data points represent the statistical and systematic uncertainties added in quadrature.
    The inner section of the vertical bars, denoted by the tick marks, show the statistical uncertainty.
    Left: comparison with different
    simulation event generators: \MADGRAPHPYTHIA (both the default and after reweighting the top quark \pt spectrum),
    \POWHEGVTWOHERWIG, and \POWHEGVTWOPYTHIA. Right: comparison with
    predictions from the \MADGRAPHPYTHIA event generator found by varying the matching threshold and renormalization
    scales ($\mu_{\mathrm{R}}$, $\mu_{\mathrm{F}}$) up and down by a factor of two.  The lower plots show the ratio of the
    predictions to the data, with the statistical and total uncertainties in the ratios indicated by the two shaded bands.}
    \label{fig:ST_WPT_result_combined_7TeV}
  \end{center}
\end{figure*}

\begin{figure*}[!htb]
  \begin{center}
      \includegraphics[width=\cmsFigWidth]{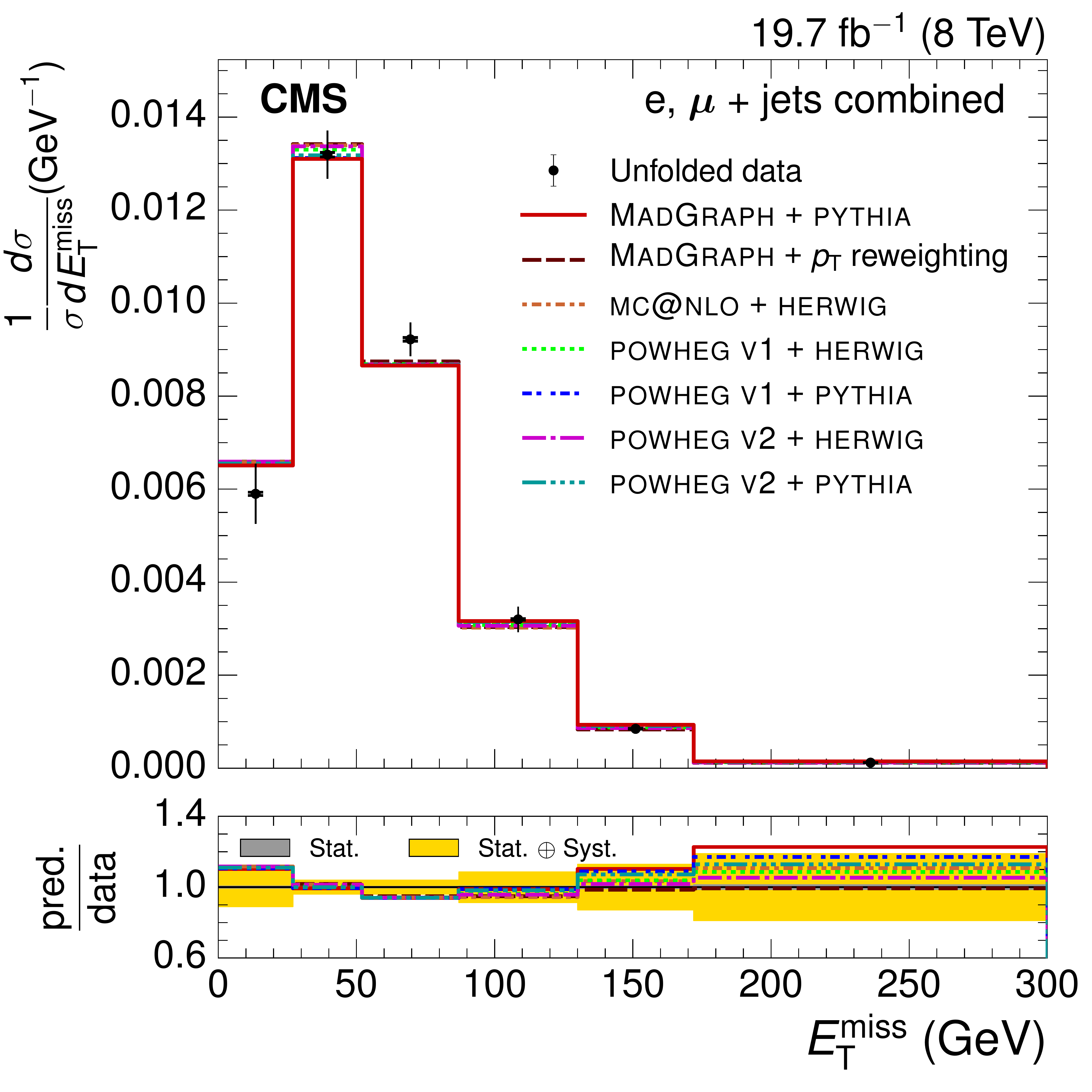}\hfill
      \includegraphics[width=\cmsFigWidth]{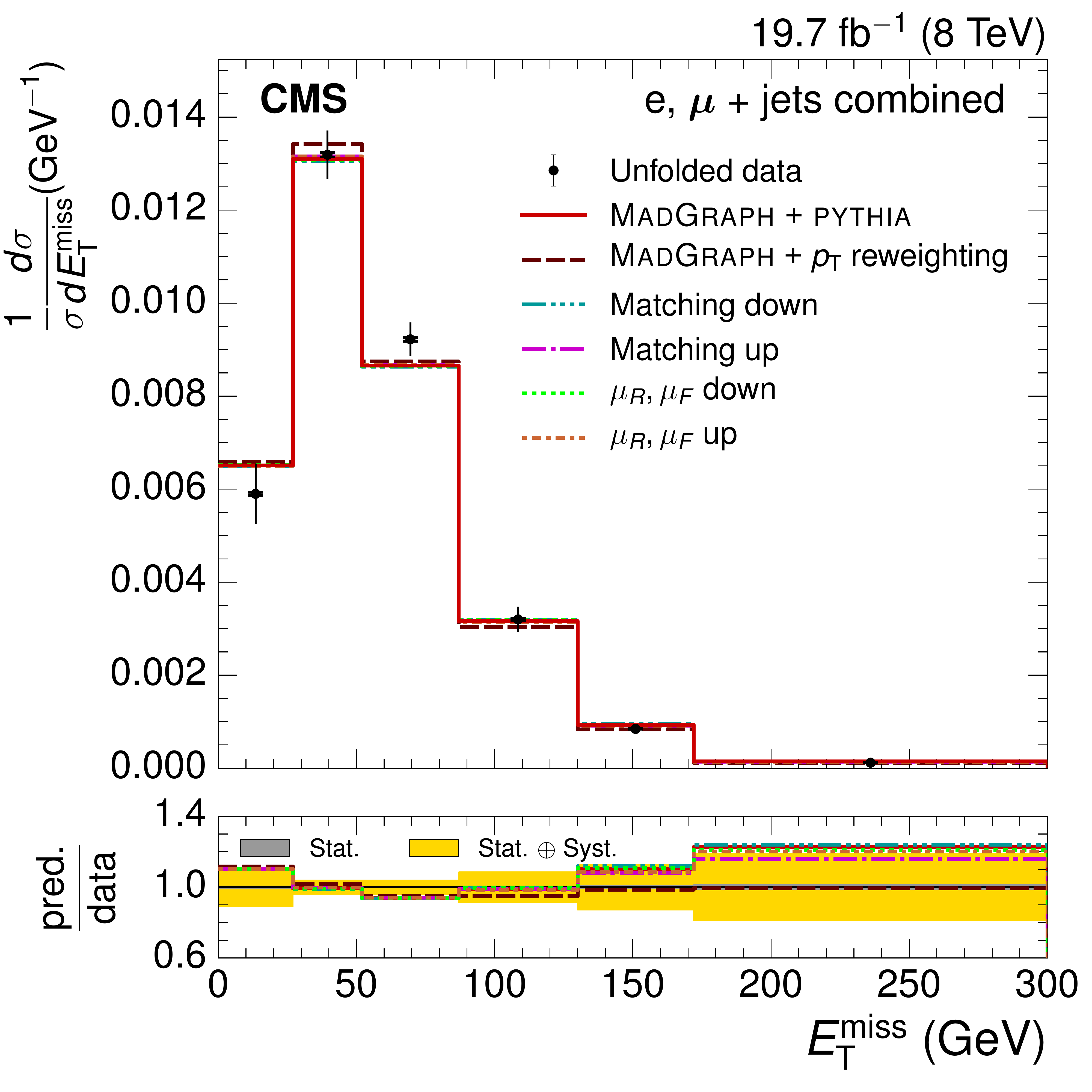} \\
      \includegraphics[width=\cmsFigWidth]{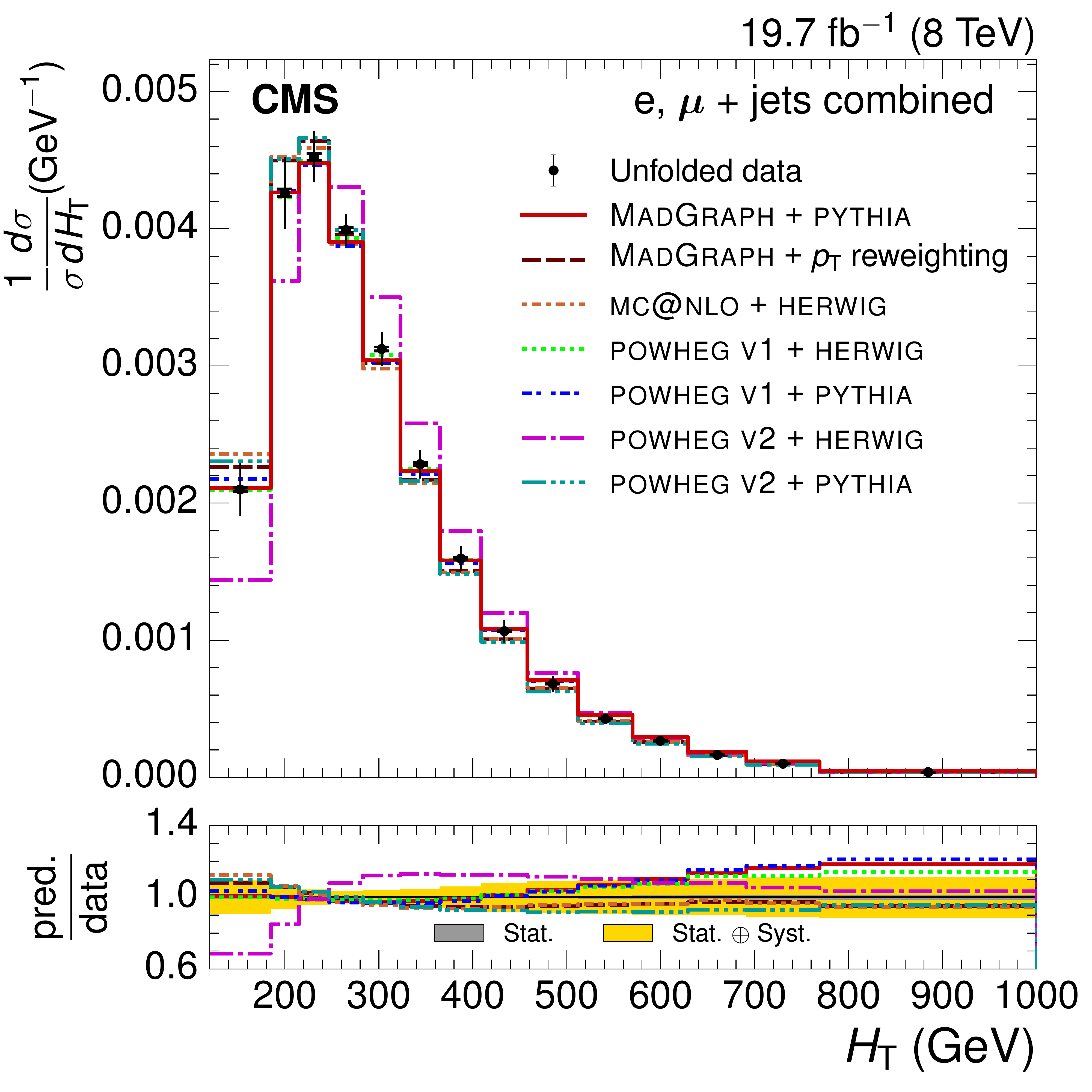}\hfill
      \includegraphics[width=\cmsFigWidth]{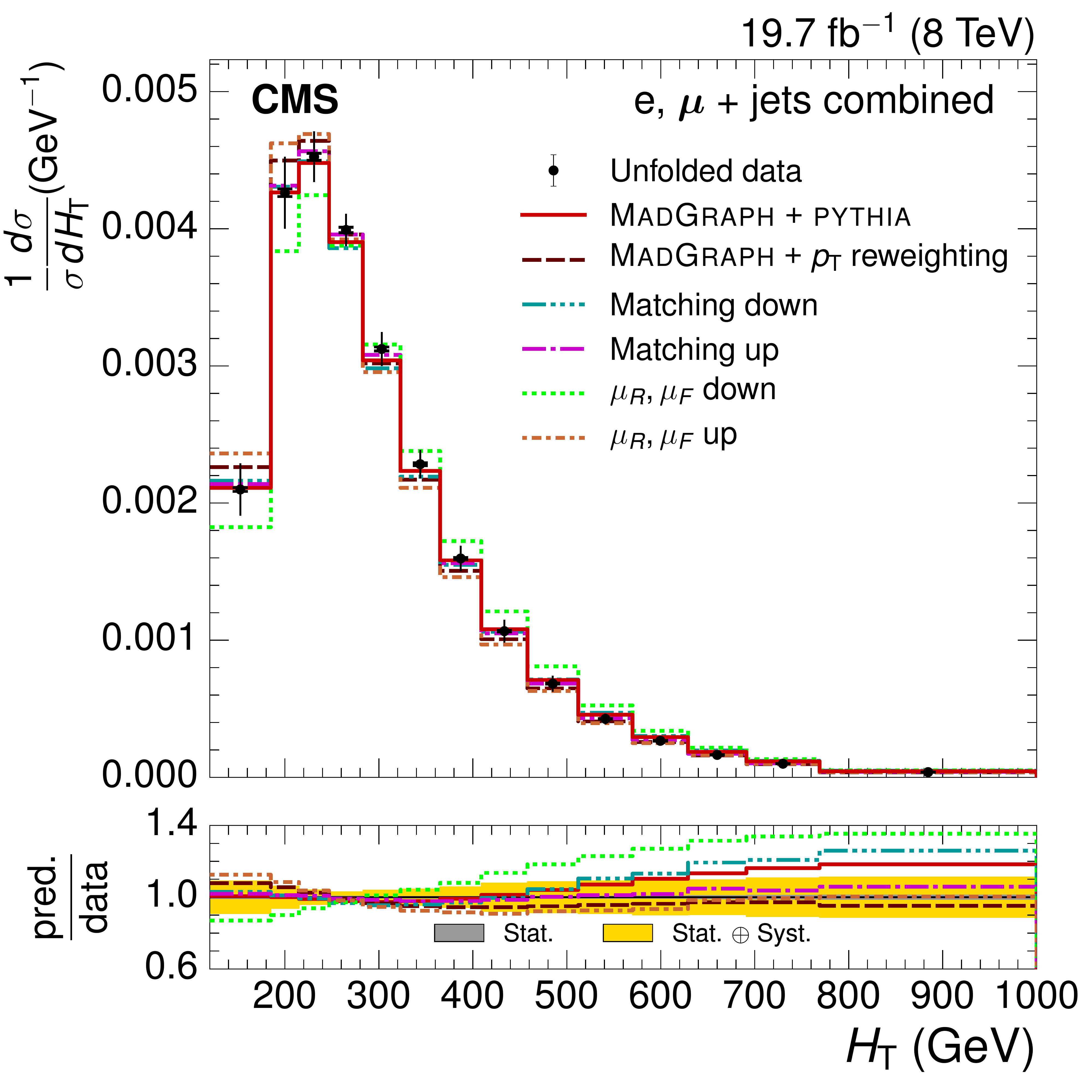}
    \caption{Normalized \met (top) and \HT (bottom) differential \ttbar cross sections from combined
    electron and muon data at $\sqrt{s}=8\TeV$.
    The vertical bars on the data points represent the statistical and systematic uncertainties added in quadrature.
    The inner section of the vertical bars, denoted by the tick marks, show the statistical uncertainty.
    Left: comparison with different
    simulation event generators: \MADGRAPHPYTHIA (both the default and after reweighting the top quark \pt spectrum),
    \MCATNLOHERWIG, \POWHEGVONEHERWIG, \POWHEGVONEPYTHIA, \POWHEGVTWOHERWIG, and \POWHEGVTWOPYTHIA.
    Right: comparison with
    predictions from the \PYTHIA event generator found by varying the matching threshold and renormalization
    scales ($\mu_{\mathrm{R}}$, $\mu_{\mathrm{F}}$) up and down by a factor of two.  The lower plots show the ratio of the
    predictions to the data, with the statistical and total uncertainties in the ratios indicated by the two shaded bands.}
      \label{fig:MET_HT_result_combined_8TeV}
  \end{center}
\end{figure*}

\begin{figure*}[!htb]
  \begin{center}
      \includegraphics[width=\cmsFigWidth]{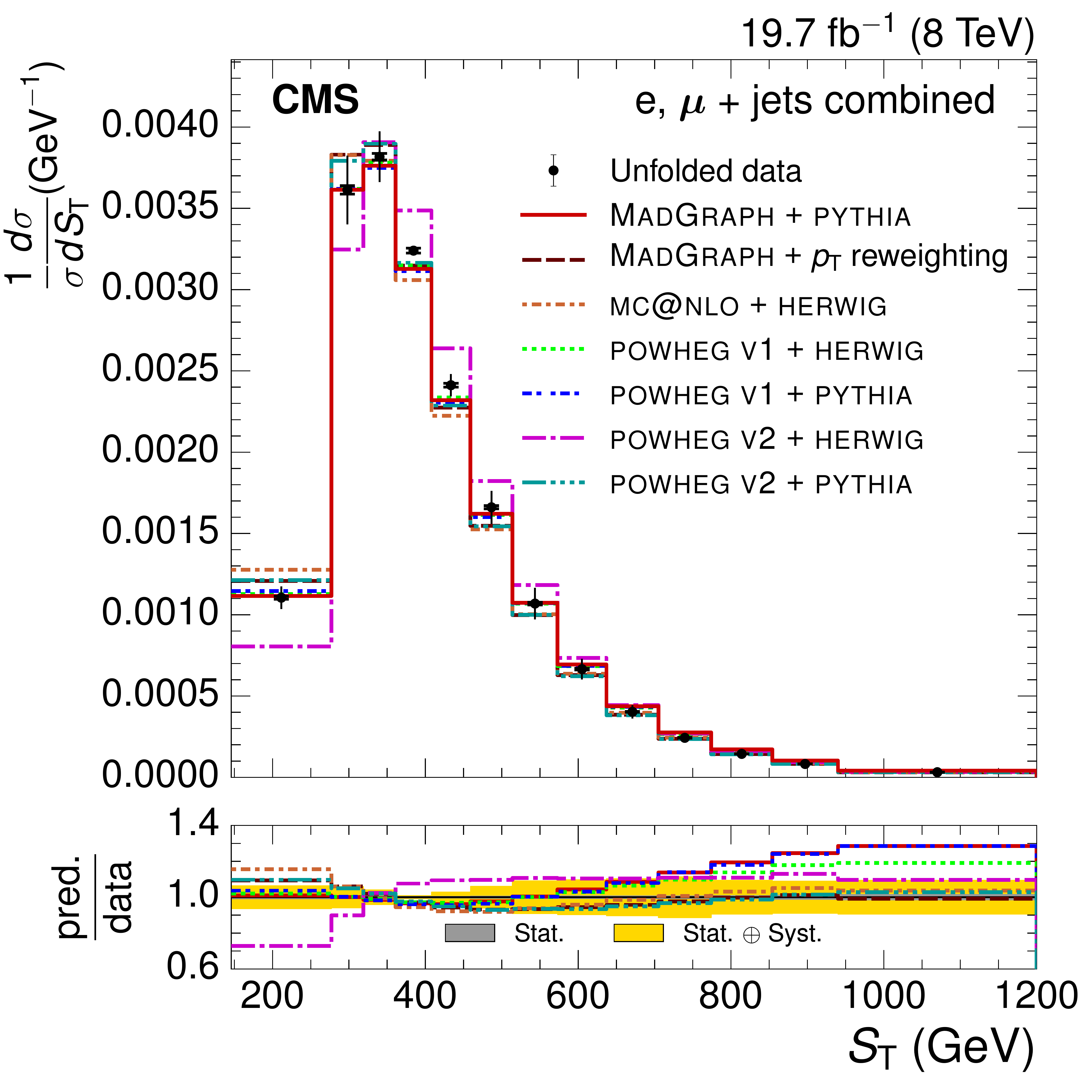}\hfill
      \includegraphics[width=\cmsFigWidth]{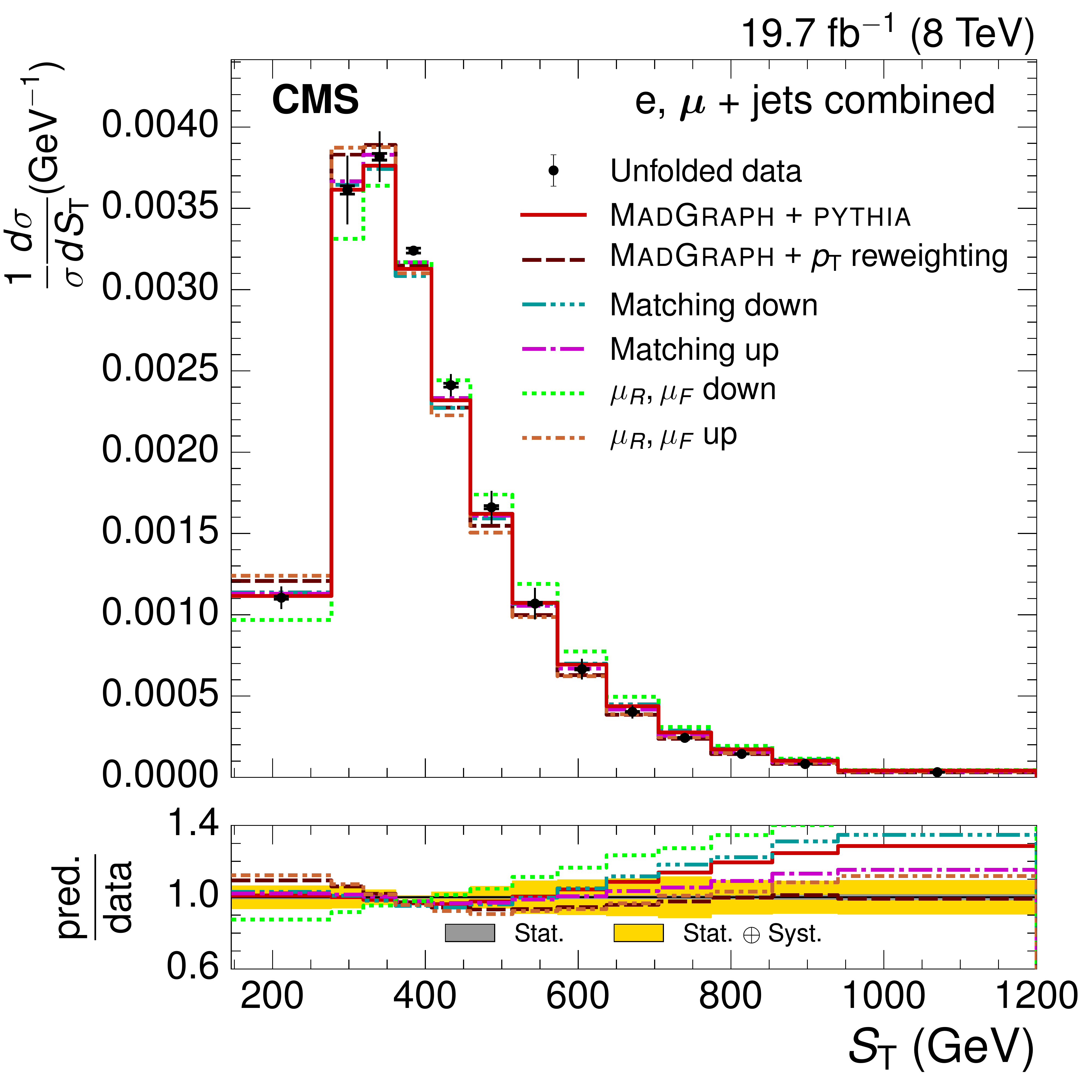} \\
      \includegraphics[width=\cmsFigWidth]{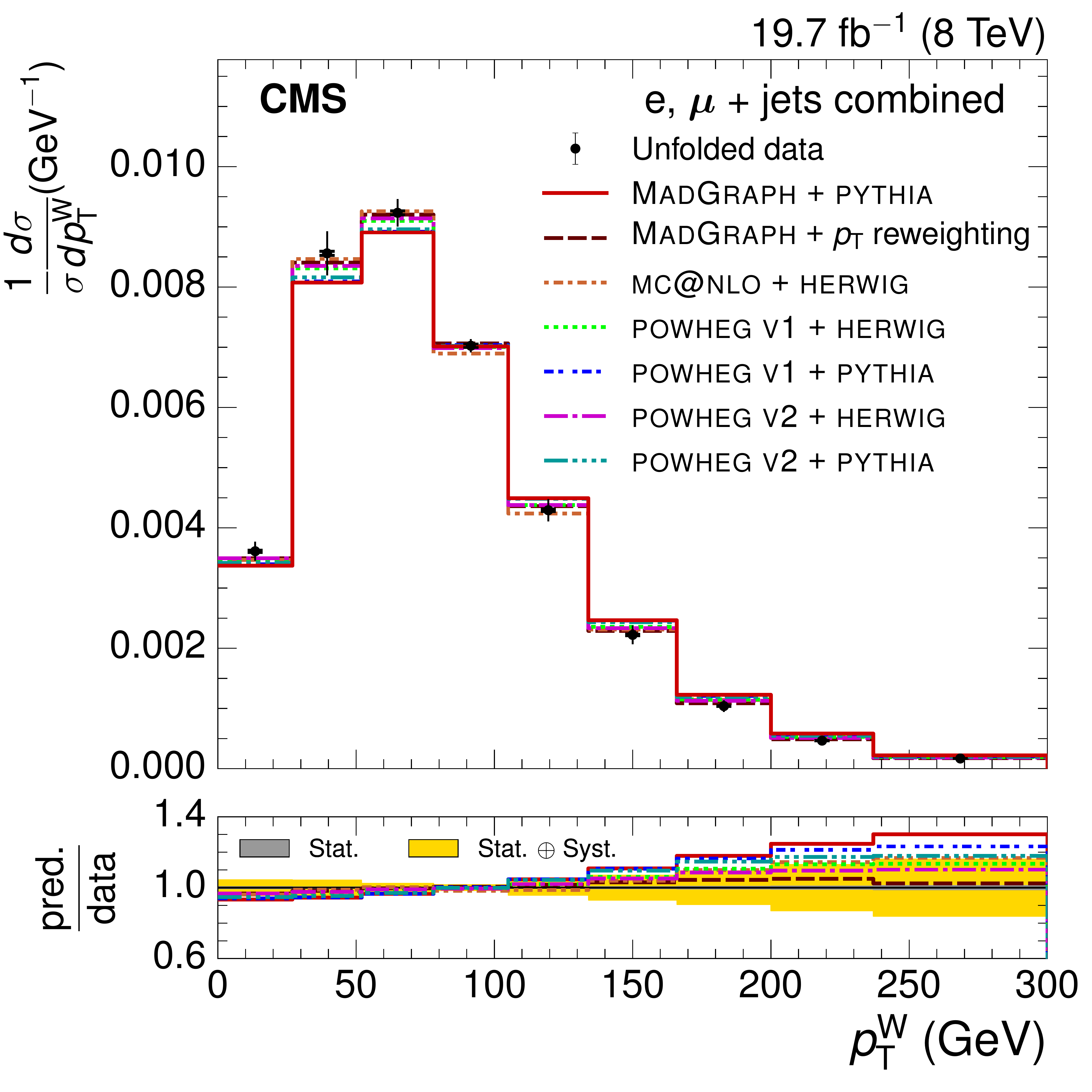}\hfill
      \includegraphics[width=\cmsFigWidth]{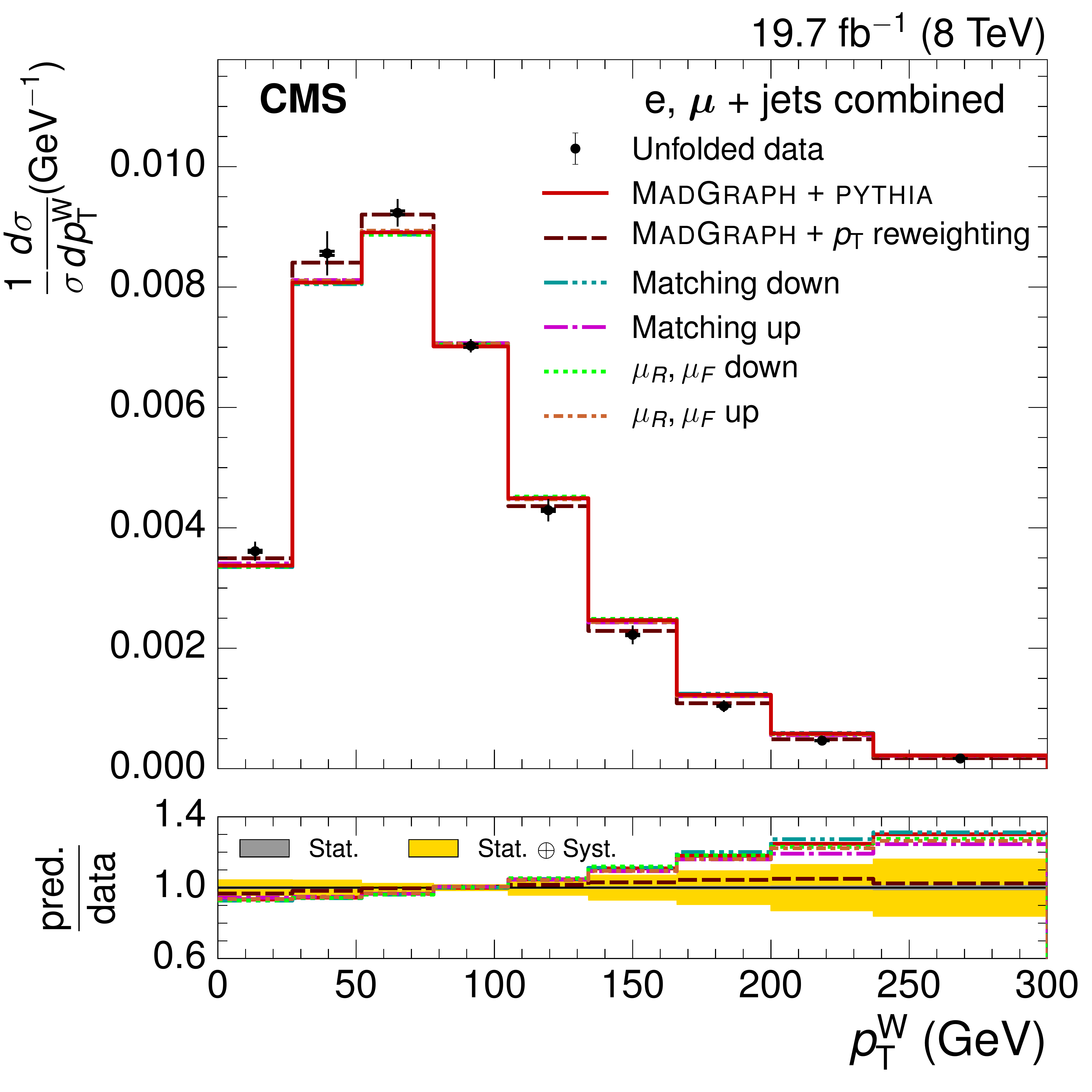}
    \caption{Normalized \ST (top) and \wpt (bottom) differential \ttbar cross sections from combined
    electron and muon data at $\sqrt{s}=8\TeV$.
    The vertical bars on the data points represent the statistical and systematic uncertainties added in quadrature.
    The inner section of the vertical bars, denoted by the tick marks, show the statistical uncertainty.
    Left: comparison with different
    simulation event generators: \MADGRAPHPYTHIA (both the default and after reweighting the top quark \pt spectrum),
    \MCATNLOHERWIG, \POWHEGVONEHERWIG, \POWHEGVONEPYTHIA, \POWHEGVTWOHERWIG, and \POWHEGVTWOPYTHIA.
    Right: comparison with
    predictions from the \MADGRAPHPYTHIA event generator found by varying the matching threshold and renormalization
    scales ($\mu_{\mathrm{R}}$, $\mu_{\mathrm{F}}$) up and down by a factor of two.  The lower plots show the ratio of the
    predictions to the data, with the statistical and total uncertainties in the ratios indicated by the two shaded bands.}
      \label{fig:ST_WPT_result_combined_8TeV}
  \end{center}
\end{figure*}

The normalized differential \ttbar cross sections as a function of each of the kinematic event variables are shown in
Figs.~\ref{fig:MET_HT_result_combined_7TeV} and \ref{fig:ST_WPT_result_combined_7TeV} for the \sqrtSeven data, and in
Figs.~\ref{fig:MET_HT_result_combined_8TeV} and \ref{fig:ST_WPT_result_combined_8TeV} for the \sqrtEight data.
The results are also presented in \suppMaterial.

The data distributions in the figures are compared with the predictions
from the event generators in the left-hand plots: \MADGRAPH
and \POWHEGVTWO with two different hadron shower generators, \PYTHIA and \HERWIG.
For the \sqrtEight results, the predictions from the \MCATNLO and \POWHEGVONE generators are also shown.
The effect on the predicted distributions from varying the modeling parameters (the matching threshold and
renormalization scale $Q^2$) up and down by a factor of two for the \MADGRAPH event generator is shown in the
right-hand plots for the two \MADGRAPH simulations. The uncertainties shown by the vertical bars
on the points in the figures and given in the tables
include both the statistical uncertainties and those resulting from the unfolding procedure.

The measurements at \sqrtSeven are well described by all the event generators
in the distribution of \met.  For \st, \wpt, and \HT,
the event generators predict a somewhat harder spectrum than seen in data.
However, the \POWHEGVTWOPYTHIA event generator
provides a reasonable description of the \HT and \st differential cross sections.

The results at \sqrtEight are generally well described by the \MCATNLO and the
\POWHEGVTWOPYTHIA event generators.
The \POWHEGVTWOHERWIG event generator describes the \met and \wpt distributions well.
However, for \HT and \st~
this event generator predicts a harder spectrum than seen in data, at both center-of-mass energies.

The \MADGRAPH event generator generally predicts a harder spectrum than seen in data for all variables.
The variations in matching threshold and $Q^2$ in the \MADGRAPH event generator are not sufficient to
explain this difference between the prediction and data.
However, the \MADGRAPH event generator provides a good description of the
data after reweighting the top quark \pt spectrum, as described in Section~\ref{sec:mc_modelling}.  The prediction
obtained from the \MADGRAPH event generator after the reweighting is shown on all the plots.

\section{Summary}

A measurement of the normalized differential cross section of top quark pair production with respect to the four
kinematic event variables \met, \HT, \st,\ and \wpt\
has been performed in $\Pp\Pp$ collisions at a center-of-mass energy of 7\TeV using 5.0\fbinv and at 8\TeV using
19.7\fbinv of data collected by the CMS experiment.

This study confirms previous CMS findings that the observed
top quark \pt spectrum is softer than predicted
by the \MADGRAPH, \POWHEG, and
\MCATNLO event generators, but otherwise there is broad consistency
between the MC event generators and observation.
This result provides confidence in the description
of \ttbar production in the SM and its implementation
in the most frequently used simulation packages.

\begin{acknowledgments}

We congratulate our colleagues in the CERN accelerator departments for the excellent performance of the LHC and thank the technical and administrative staffs at CERN and at other CMS institutes for their contributions to the success of the CMS effort. In addition, we gratefully acknowledge the computing centers and personnel of the Worldwide LHC Computing Grid for delivering so effectively the computing infrastructure essential to our analyses. Finally, we acknowledge the enduring support for the construction and operation of the LHC and the CMS detector provided by the following funding agencies: BMWFW and FWF (Austria); FNRS and FWO (Belgium); CNPq, CAPES, FAPERJ, and FAPESP (Brazil); MES (Bulgaria); CERN; CAS, MoST, and NSFC (China); COLCIENCIAS (Colombia); MSES and CSF (Croatia); RPF (Cyprus); SENESCYT (Ecuador); MoER, ERC IUT and ERDF (Estonia); Academy of Finland, MEC, and HIP (Finland); CEA and CNRS/IN2P3 (France); BMBF, DFG, and HGF (Germany); GSRT (Greece); OTKA and NIH (Hungary); DAE and DST (India); IPM (Iran); SFI (Ireland); INFN (Italy); MSIP and NRF (Republic of Korea); LAS (Lithuania); MOE and UM (Malaysia); BUAP, CINVESTAV, CONACYT, LNS, SEP, and UASLP-FAI (Mexico); MBIE (New Zealand); PAEC (Pakistan); MSHE and NSC (Poland); FCT (Portugal); JINR (Dubna); MON, RosAtom, RAS and RFBR (Russia); MESTD (Serbia); SEIDI and CPAN (Spain); Swiss Funding Agencies (Switzerland); MST (Taipei); ThEPCenter, IPST, STAR and NSTDA (Thailand); TUBITAK and TAEK (Turkey); NASU and SFFR (Ukraine); STFC (United Kingdom); DOE and NSF (USA).

Individuals have received support from the Marie-Curie program and the European Research Council and EPLANET (European Union); the Leventis Foundation; the A. P. Sloan Foundation; the Alexander von Humboldt Foundation; the Belgian Federal Science Policy Office; the Fonds pour la Formation \`a la Recherche dans l'Industrie et dans l'Agriculture (FRIA-Belgium); the Agentschap voor Innovatie door Wetenschap en Technologie (IWT-Belgium); the Ministry of Education, Youth and Sports (MEYS) of the Czech Republic; the Council of Science and Industrial Research, India; the HOMING PLUS program of the Foundation for Polish Science, cofinanced from European Union, Regional Development Fund, the Mobility Plus program of the Ministry of Science and Higher Education, the OPUS program contract 2014/13/B/ST2/02543 and contract Sonata-bis DEC-2012/07/E/ST2/01406 of the National Science Center (Poland); the Thalis and Aristeia programmes cofinanced by EU-ESF and the Greek NSRF; the National Priorities Research Program by Qatar National Research Fund; the Programa Clar\'in-COFUND del Principado de Asturias; the Rachadapisek Sompot Fund for Postdoctoral Fellowship, Chulalongkorn University and the Chulalongkorn Academic into Its 2nd Century Project Advancement Project (Thailand); and the Welch Foundation, contract C-1845.

\end{acknowledgments}

\bibliography{auto_generated}

\numberwithin{figure}{section}
\numberwithin{table}{section}
\appendix
\clearpage
\section{\label{s:additional_tables}Additional tables}

The measured values of the \ttbar differential cross sections as a function of \met, \HT, \st, and \wpt
for \sqrtSeven and \sqrtEight are given in the tables below, along with their statistical, systematic, and total
uncertainties.

\begin{table}[htbp]
\centering
\topcaption{Normalized \ttbar differential cross section measurements with respect to the \met variable
at a center-of-mass energy of 7\TeV (combination of electron and muon channels). The rightmost three columns show the relative uncertainties on the measured values, in percent. The statistical and systematic uncertainties are listed separately, and are combined in quadrature to give the overall relative uncertainty.}
\label{tab:MET_xsections_7TeV_combined}
\begin{scotch}{ccccc}
\met & $1/\sigma\ \rd \sigma/\rd \met$ &  $\pm \text{ stat.}$ & $\pm \text{ syst.}$ & Rel. uncert. \\
(\GeVns{}) & ($\GeVns^{-1}$) & (\%) & (\%) & (\%)\\
\hline
\x0--27 &  $6.44 \times 10^{-3}$ & $0.83$ & $4.5$ & $4.6$\\
27--52 &  $1.32 \times 10^{-2}$ & $0.60$ & $2.8$ & $2.9$\\
52--87 &  $8.75 \times 10^{-3}$ & $0.58$ & $1.9$ & $2.0$\\
\x87--130 &  $3.14 \times 10^{-3}$ & $0.80$ & $6.0$ & $6.0$\\
130--172 &  $8.93 \times 10^{-4}$ & $1.1$ & $12$ & $12$\\
172--300 &  $1.32 \times 10^{-4}$ & $1.4$ & $19$ & $19$\\
\end{scotch}
\end{table}

\begin{table}[htbp]
\centering
\topcaption{Normalized \ttbar differential cross section measurements with respect to the \HT variable
at a center-of-mass energy of 7 TeV (combination of electron and muon channels). The rightmost three columns show the relative uncertainties on the measured values, in percent. The statistical and systematic uncertainties are listed separately, and are combined in quadrature to give the overall relative uncertainty.}
\label{tab:HT_xsections_7TeV_combined}
\begin{scotch}{ccccc}
\HT & $1/\sigma\ \rd \sigma/\rd \HT$ &  $\pm \text{ stat.}$ & $\pm \text{ syst.}$ & Rel. uncert. \\
(\GeVns{}) & ($\GeVns^{-1}$) & (\%) & (\%) & (\%)\\
\hline
120--185 &  $2.48 \times 10^{-3}$ & $1.5$ & $6.8$ & $6.9$\\
185--215 &  $4.86 \times 10^{-3}$ & $1.4$ & $5.5$ & $5.7$\\
215--247 &  $4.89 \times 10^{-3}$ & $1.3$ & $4.4$ & $4.6$\\
247--283 &  $4.05 \times 10^{-3}$ & $1.2$ & $2.8$ & $3.1$\\
283--323 &  $2.99 \times 10^{-3}$ & $1.1$ & $2.9$ & $3.1$\\
323--365 &  $2.06 \times 10^{-3}$ & $1.1$ & $5.4$ & $5.6$\\
365--409 &  $1.37 \times 10^{-3}$ & $1.1$ & $7.0$ & $7.1$\\
409--458 &  $8.93 \times 10^{-4}$ & $1.1$ & $9.3$ & $9.4$\\
458--512 &  $5.49 \times 10^{-4}$ & $1.2$ & $9.9$ & $10$\\
512--570 &  $3.38 \times 10^{-4}$ & $1.4$ & $13$ & $13$\\
570--629 &  $2.04 \times 10^{-4}$ & $1.8$ & $10$ & $11$\\
629--691 &  $1.25 \times 10^{-4}$ & $2.2$ & $14$ & $14$\\
691--769 &  $7.20 \times 10^{-5}$ & $2.7$ & $12$ & $13$\\
\x769--1000 &  $2.51 \times 10^{-5}$ & $3.0$ & $17$ & $17$\\
\end{scotch}
\end{table}

\begin{table}[htbp]
\centering
\topcaption{Normalized \ttbar differential cross section measurements with respect to the \st variable
at a center-of-mass energy of 7 TeV (combination of electron and muon channels). The rightmost three columns show the relative uncertainties on the measured values, in percent. The statistical and systematic uncertainties are listed separately, and are combined in quadrature to give the overall relative uncertainty.}
\label{tab:ST_xsections_7TeV_combined}
\begin{scotch}{ccccc}
\st & $1/\sigma\ \rd \sigma/\rd \st$ &  $\pm \text{ stat.}$ & $\pm \text{ syst.}$ & Rel. uncert. \\
(\GeVns{}) & ($\GeVns^{-1}$) & (\%) & (\%) & (\%)\\
\hline
146--277 &  $1.31 \times 10^{-3}$ & $1.2$ & $8.4$ & $8.5$\\
277--319 &  $4.12 \times 10^{-3}$ & $1.1$ & $6.7$ & $6.8$\\
319--361 &  $4.05 \times 10^{-3}$ & $1.0$ & $4.2$ & $4.3$\\
361--408 &  $3.18 \times 10^{-3}$ & $0.91$ & $1.8$ & $2.0$\\
408--459 &  $2.21 \times 10^{-3}$ & $0.93$ & $4.5$ & $4.6$\\
459--514 &  $1.44 \times 10^{-3}$ & $1.0$ & $8.1$ & $8.2$\\
514--573 &  $8.96 \times 10^{-4}$ & $1.1$ & $10$ & $11$\\
573--637 &  $5.42 \times 10^{-4}$ & $1.2$ & $11$ & $11$\\
637--705 &  $3.25 \times 10^{-4}$ & $1.3$ & $11$ & $11$\\
705--774 &  $1.95 \times 10^{-4}$ & $1.6$ & $12$ & $13$\\
774--854 &  $1.13 \times 10^{-4}$ & $1.9$ & $12$ & $12$\\
854--940 &  $6.32 \times 10^{-5}$ & $2.3$ & $10$ & $10$\\
\x940--1200 &  $2.26 \times 10^{-5}$ & $2.7$ & $14$ & $14$\\
\end{scotch}
\end{table}

\begin{table}[htbp]
\centering
\topcaption{Normalized \ttbar differential cross section measurements with respect to the \wpt variable
at a center-of-mass energy of 7 TeV (combination of electron and muon channels). The rightmost three columns show the relative uncertainties on the measured values, in percent. The statistical and systematic uncertainties are listed separately, and are combined in quadrature to give the overall relative uncertainty.}
\label{tab:WPT_xsections_7TeV_combined}
\begin{scotch}{ccccc}
\wpt & $1/\sigma\ \rd \sigma/\rd \wpt$ &  $\pm \text{ stat.}$ & $\pm \text{ syst.}$ & Rel. uncert. \\
(\GeVns{}) & ($\GeVns^{-1}$) & (\%) & (\%) & (\%)\\
\hline
\x0--27 &  $3.58 \times 10^{-3}$ & $1.3$ & $3.8$ & $4.1$\\
27--52 &  $8.56 \times 10^{-3}$ & $0.96$ & $3.4$ & $3.6$\\
52--78 &  $9.33 \times 10^{-3}$ & $0.81$ & $2.5$ & $2.6$\\
78--105 &  $7.06 \times 10^{-3}$ & $0.96$ & $1.9$ & $2.1$\\
105--134 &  $4.28 \times 10^{-3}$ & $1.2$ & $4.1$ & $4.2$\\
134--166 &  $2.20 \times 10^{-3}$ & $1.3$ & $6.1$ & $6.2$\\
166--200 &  $1.02 \times 10^{-3}$ & $1.6$ & $8.0$ & $8.1$\\
200--237 &  $4.56 \times 10^{-4}$ & $2.2$ & $9.9$ & $10$\\
237--300 &  $1.63 \times 10^{-4}$ & $2.9$ & $13$ & $13$\\
\end{scotch}
\end{table}

\begin{table}[htbp]
\centering
\topcaption{Normalized \ttbar differential cross section measurements with respect to the \met variable
at a center-of-mass energy of 8 TeV (combination of electron and muon channels). The rightmost three columns show the relative uncertainties on the measured values, in percent. The statistical and systematic uncertainties are listed separately, and are combined in quadrature to give the overall relative uncertainty.}
\label{tab:MET_xsections_8TeV_combined}
\begin{scotch}{ccccc}
\met & $1/\sigma\ \rd \sigma/\rd \met$ &  $\pm \text{ stat.}$ & $\pm \text{ syst.}$ & Rel. uncert. \\
(\GeVns{}) & ($\GeVns^{-1}$) & (\%) & (\%) & (\%)\\
\hline
\x0--27 &  $5.90 \times 10^{-3}$ & $0.59$ & $11$ & $11$\\
27--52 &  $1.32 \times 10^{-2}$ & $0.36$ & $3.9$ & $3.9$\\
52--87 &  $9.22 \times 10^{-3}$ & $0.40$ & $3.9$ & $3.9$\\
\x87--130 &  $3.20 \times 10^{-3}$ & $0.55$ & $8.6$ & $8.7$\\
130--172 &  $8.46 \times 10^{-4}$ & $0.81$ & $13$ & $13$\\
172--300 &  $1.18 \times 10^{-4}$ & $1.3$ & $19$ & $19$\\
\end{scotch}
\end{table}

\begin{table}[htbp]
\centering
\topcaption{Normalized \ttbar differential cross section measurements with respect to the \HT variable
at a center-of-mass energy of 8 TeV (combination of electron and muon channels). The rightmost three columns show the relative uncertainties on the measured values, in percent. The statistical and systematic uncertainties are listed separately, and are combined in quadrature to give the overall relative uncertainty.}
\label{tab:HT_xsections_8TeV_combined}
\begin{scotch}{ccccc}
\HT & $1/\sigma\ \rd \sigma/\rd \HT$ &  $\pm \text{ stat.}$ & $\pm \text{ syst.}$ & Rel. uncert. \\
(\GeVns{}) & ($\GeVns^{-1}$) & (\%) & (\%) & (\%)\\
\hline
120--185 &  $2.10 \times 10^{-3}$ & $0.68$ & $9.1$ & $9.1$\\
185--215 &  $4.26 \times 10^{-3}$ & $0.65$ & $6.1$ & $6.2$\\
215--247 &  $4.52 \times 10^{-3}$ & $0.57$ & $4.1$ & $4.1$\\
247--283 &  $3.99 \times 10^{-3}$ & $0.50$ & $2.9$ & $3.0$\\
283--323 &  $3.12 \times 10^{-3}$ & $0.46$ & $4.0$ & $4.0$\\
323--365 &  $2.28 \times 10^{-3}$ & $0.44$ & $4.5$ & $4.6$\\
365--409 &  $1.60 \times 10^{-3}$ & $0.44$ & $5.8$ & $5.8$\\
409--458 &  $1.07 \times 10^{-3}$ & $0.43$ & $7.9$ & $7.9$\\
458--512 &  $6.83 \times 10^{-4}$ & $0.45$ & $8.6$ & $8.6$\\
512--570 &  $4.26 \times 10^{-4}$ & $0.51$ & $9.0$ & $9.0$\\
570--629 &  $2.66 \times 10^{-4}$ & $0.65$ & $9.9$ & $9.9$\\
629--691 &  $1.64 \times 10^{-4}$ & $0.82$ & $9.7$ & $9.7$\\
691--769 &  $9.93 \times 10^{-5}$ & $0.99$ & $11$ & $11$\\
\x769--1000 &  $3.78 \times 10^{-5}$ & $1.1$ & $11$ & $11$\\
\end{scotch}
\end{table}

\begin{table}[htbp]
\centering
\topcaption{Normalized \ttbar differential cross section measurements with respect to the \st variable
at a center-of-mass energy of 8 TeV (combination of electron and muon channels). The rightmost three columns show the relative uncertainties on the measured values, in percent. The statistical and systematic uncertainties are listed separately, and are combined in quadrature to give the overall relative uncertainty.}
\label{tab:ST_xsections_8TeV_combined}
\begin{scotch}{ccccc}
\st & $1/\sigma\ \rd \sigma/\rd \st$ &  $\pm \text{ stat.}$ & $\pm \text{ syst.}$ & Rel. uncert. \\
(\GeVns{}) & ($\GeVns^{-1}$) & (\%) & (\%) & (\%)\\
\hline
146--277 &  $1.10 \times 10^{-3}$ & $0.84$ & $6.3$ & $6.3$\\
277--319 &  $3.61 \times 10^{-3}$ & $0.71$ & $5.8$ & $5.9$\\
319--361 &  $3.82 \times 10^{-3}$ & $0.54$ & $4.1$ & $4.1$\\
361--408 &  $3.24 \times 10^{-3}$ & $0.46$ & $0.80$ & $0.92$\\
408--459 &  $2.41 \times 10^{-3}$ & $0.48$ & $2.8$ & $2.9$\\
459--514 &  $1.66 \times 10^{-3}$ & $0.57$ & $6.1$ & $6.1$\\
514--573 &  $1.07 \times 10^{-3}$ & $0.69$ & $9.0$ & $9.1$\\
573--637 &  $6.65 \times 10^{-4}$ & $0.74$ & $9.6$ & $9.6$\\
637--705 &  $4.03 \times 10^{-4}$ & $0.71$ & $10$ & $10$\\
705--774 &  $2.43 \times 10^{-4}$ & $0.73$ & $11$ & $11$\\
774--854 &  $1.44 \times 10^{-4}$ & $0.88$ & $9.3$ & $9.4$\\
854--940 &  $8.21 \times 10^{-5}$ & $1.2$ & $8.9$ & $9.0$\\
\x940--1200 &  $3.15 \times 10^{-5}$ & $1.5$ & $9.2$ & $9.4$\\
\end{scotch}
\end{table}

\begin{table}[htbp]
\centering
\topcaption{Normalized \ttbar differential cross section measurements with respect to the \wpt variable
at a center-of-mass energy of 8 TeV (combination of electron and muon channels). The rightmost three columns show the relative uncertainties on the measured values, in percent. The statistical and systematic uncertainties are listed separately, and are combined in quadrature to give the overall relative uncertainty.}
\label{tab:WPT_xsections_8TeV_combined}
\begin{scotch}{ccccc}
\wpt & $1/\sigma\ \rd \sigma/\rd \wpt$ &  $\pm \text{ stat.}$ & $\pm \text{ syst.}$ & Rel. uncert.  \\
(\GeVns{}) & ($\GeVns^{-1}$) & (\%) & (\%) & (\%)\\
\hline
\x0--27 &  $3.61 \times 10^{-3}$ & $0.54$ & $4.4$ & $4.4$\\
27--52 &  $8.56 \times 10^{-3}$ & $0.40$ & $4.3$ & $4.3$\\
52--78 &  $9.23 \times 10^{-3}$ & $0.34$ & $2.5$ & $2.5$\\
\x78--105 &  $7.02 \times 10^{-3}$ & $0.40$ & $1.6$ & $1.6$\\
105--134 &  $4.29 \times 10^{-3}$ & $0.50$ & $4.3$ & $4.3$\\
134--166 &  $2.22 \times 10^{-3}$ & $0.55$ & $7.1$ & $7.1$\\
166--200 &  $1.04 \times 10^{-3}$ & $0.67$ & $9.6$ & $9.6$\\
200--237 &  $4.66 \times 10^{-4}$ & $0.94$ & $13$ & $13$\\
237--300 &  $1.69 \times 10^{-4}$ & $1.2$ & $16$ & $16$\\
\end{scotch}
\end{table}

\clearpage

\cleardoublepage \section{The CMS Collaboration \label{app:collab}}\begin{sloppypar}\hyphenpenalty=5000\widowpenalty=500\clubpenalty=5000\textbf{Yerevan Physics Institute,  Yerevan,  Armenia}\\*[0pt]
V.~Khachatryan, A.M.~Sirunyan, A.~Tumasyan
\vskip\cmsinstskip
\textbf{Institut f\"{u}r Hochenergiephysik der OeAW,  Wien,  Austria}\\*[0pt]
W.~Adam, E.~Asilar, T.~Bergauer, J.~Brandstetter, E.~Brondolin, M.~Dragicevic, J.~Er\"{o}, M.~Flechl, M.~Friedl, R.~Fr\"{u}hwirth\cmsAuthorMark{1}, V.M.~Ghete, C.~Hartl, N.~H\"{o}rmann, J.~Hrubec, M.~Jeitler\cmsAuthorMark{1}, V.~Kn\"{u}nz, A.~K\"{o}nig, M.~Krammer\cmsAuthorMark{1}, I.~Kr\"{a}tschmer, D.~Liko, T.~Matsushita, I.~Mikulec, D.~Rabady\cmsAuthorMark{2}, B.~Rahbaran, H.~Rohringer, J.~Schieck\cmsAuthorMark{1}, R.~Sch\"{o}fbeck, J.~Strauss, W.~Treberer-Treberspurg, W.~Waltenberger, C.-E.~Wulz\cmsAuthorMark{1}
% ...
\vskip\cmsinstskip
% ...
1:~~Also at Vienna University of Technology, Vienna, Austria\\
2:~~Also at CERN, European Organization for Nuclear Research, Geneva, Switzerland\\
% ...

\end{sloppypar}
\end{document}